\documentclass[aps,prl,twocolumn,notitlepage]{revtex4-2}
\usepackage{epsfig,amssymb,amsmath,mathrsfs,amsfonts,amsthm,graphicx,float,color}
 \def\map#1{\mathcal #1}
\def\d{\operatorname{d}}\def\<{\langle}\def\>{\rangle}

\def\Tr{\operatorname{Tr}}\def\:{\hbox{\bf
    :}}

\def\R{\mathbb R}

\def\C{\mathbb C}
\def\grp#1{\mathsf{#1}}
\def\st#1{\mathbf{#1}}

\def\spc#1{\mathcal{#1}}

\def\set#1{\mathsf{#1}}

\def\Proof{{\bf Proof.}}

\newtheorem{theo}{{Theorem}}
\newtheorem{lem}{{Lemma}}
\newtheorem{prop}{{Proposition}}
\newtheorem{cor}{{Corollary}}

\newtheorem{definition}{{Definition}}

\newcommand{\ket}[1]{\left|#1\right\rangle}

\newcommand{\bra}[1]{\left\langle#1\right|}

\newcommand{\ketbra}[1]{\ket{#1}\bra{#1}}

\newcommand{\Eq}[1]{Eq. (\ref{#1})}

\newcommand{\Jami}{Jamio\l{}kowski}

\begin{document}
\title{
Test one to test many: a unified approach to quantum benchmarks
  }
\author{Ge Bai}
\affiliation{Department of Computer Science, The University of Hong Kong, Pokfulam Road, Hong Kong 999077, China}
\affiliation{HKU Shenzhen Institute of Research and Innovation, Yuexing 2nd Rd Nanshan, Shenzhen 518057, China}
\author{Giulio Chiribella}
\email{giulio.chiribella@cs.ox.ac.uk}
\affiliation{Department of Computer Science, University of Oxford, Wolfson Building, Parks Road, Oxford OX1 3QD, United Kingdom}
\affiliation{CIFAR Program in Quantum Information Science, Canadian Institute for Advanced Research,
Toronto, Ontario ON M5G 1Z8, Canada}
\affiliation{Department of Computer Science, The University of Hong Kong, Pokfulam Road, Hong Kong 999077, China}
\affiliation{HKU Shenzhen Institute of Research and Innovation, Yuexing 2nd Rd Nanshan, Shenzhen 518057, China}

\begin{abstract}
Quantum benchmarks are  routinely used to validate the experimental demonstration of quantum information protocols.   Many relevant  protocols, however, involve an  infinite set of input  states, of which only a finite subset can be used to test the quality of the implementation.
This is a problem, because
 the benchmark for the finitely many states  used in the test can be  higher than the original benchmark calculated for infinitely many states. This situation arises in the teleportation and storage of coherent states, for which the benchmark  of  50\% fidelity is commonly used in experiments, although finite sets of coherent states normally lead to   higher benchmarks.
   Here we  show that  the average fidelity over all coherent states can be indirectly probed with a single setup, requiring only two-mode squeezing, a 50-50 beamsplitter, and  homodyne detection.  Our setup enables a rigorous experimental validation of  quantum teleportation, storage, amplification, attenuation,  and purification of  noisy coherent states.
  More generally, we prove that every quantum benchmark  can be tested by preparing a single entangled state and measuring a single observable.
    \end{abstract}

\maketitle


{\em Introduction.} Quantum information processing offers  compelling advantages over its classical counterpart. However,  realistic implementations  suffer from unavoidable   noise and imperfections. To demonstrate a quantum advantage,   one needs to ensure that, despite the imperfections, such  implementations achieve  performances that could not be achieved classically.

For every  given task, such as the transmission of information or its  storage  in a quantum memory, the limit that has to be surpassed in order to demonstrate a quantum advantage is called the  {\em quantum benchmark} \cite{hammerer2005quantum}.   Quantum benchmarks are routinely used in experiments of quantum teleportation \cite{furusawa1998unconditional, boschi1998experimental,sherson2006quantum,krauter2013deterministic,ren2017ground}  and in the realization of quantum memories \cite{julsgaard2004experimental,lobino2009memory,hedges2010efficient,hosseini2011unconditional}.   The theoretical values of the benchmarks  have been determined in a variety of scenarios, including the teleportation and storage of finite-dimensional quantum systems \cite{horodecki1999general,calsamiglia2009phase}, coherent states \cite{braunstein2000criteria,hammerer2005quantum}, and squeezed states \cite{adesso2008quantum,owari2008squeezing,chiribella2014quantum}.
Benchmarks for the amplification of coherent states are important for assessing the realization of deterministic \cite{pooser2009low}
as well as probabilistic \cite{kocsis2013heralded,zavatta2011high,ferreyrol2010implementation,usuga2010noise}    amplifiers, and have been theoretically studied in Refs. \cite{namiki2008fidelity,chiribella2013optimal}.
Many  benchmarks are fidelity-based, meaning that they use the fidelity \cite{uhlmann1976transition,jozsa1994fidelity} as the  figure of merit.
Other  benchmarks are entanglement-based, meaning that the figure of merit is  (a measure of) the ability  to  preserve entanglement \cite{haseler2008testing,haseler2009probing,haseler2010quantum,killoran2012quantum}.

In theory, quantum benchmarks provide rigorous criteria of quantumness. In practice, the application of these criteria can be   problematic.  The benchmarks often rank quantum devices based on their average performance on  an infinite set of input states, such as the set of all coherent states \cite{braunstein2000criteria,hammerer2005quantum,furusawa1998unconditional,chiribella2013optimal,namiki2008fidelity,kocsis2013heralded,sherson2006quantum,krauter2013deterministic,zavatta2011high,ferreyrol2010implementation,usuga2010noise,pooser2009low,julsgaard2004experimental,lobino2009memory,hedges2010efficient,hosseini2011unconditional}. In a real experiment, however, only a finite subset of inputs can be tested. The evaluation of the performance on each input requires many sessions of data collection, often amounting to a full tomography of the state \cite{zavatta2011high}.  Now, the problem is that the value of the benchmark for the finite subset of states used in the experiment   can be much larger than the theoretical benchmark. For example, the fidelity benchmark for the teleportation of uniformly distributed coherent states is 50\% \cite{braunstein2000criteria,hammerer2005quantum}, while the benchmark for just {\em two} coherent states is  at least  93.3\%, the minimum value over all pairs of coherent states \cite{namiki2008verification}.
Comparing the experimental fidelity with the theoretical benchmark requires additional assumptions on the device---{\em e.g.}, assumptions on how it {\em would} have worked if it had been tested on other inputs.     But making such assumptions is in contradiction to the purpose of  quantum benchmarks, i.e.,  to certify  quantum advantages  without  having to trust the  devices.  An alternative approach would be to perform a full tomography of the device \cite{chuang1997prescription,poyatos1997complete,leung2003choi,d2001quantum,dur2001nonlocal,altepeter2003ancilla}, but this would require a large  number of measurement settings (or even an infinite number in the case of continuous variable systems).

In this article, we   show that every quantum benchmark can be tested by a {\em single preparation setup} of an entangled  state and a {\em single measurement setup}  on the output.
More broadly, we develop a unified framework for quantum benchmarks, including fidelity-based and entanglement-based benchmarks as special cases. We observe that the same benchmark can be tested in  multiple equivalent ways,  among which one can choose the most experimentally friendly one.
Using the idea of equivalent tests, we propose a benchmark setup  for the demonstration of continuous-variable quantum memories  \cite{julsgaard2004experimental,lobino2009memory,hedges2010efficient,hosseini2011unconditional}  and for the demonstration of  quantum-enhanced amplification   \cite{pooser2009low,kocsis2013heralded,zavatta2011high,ferreyrol2010implementation,usuga2010noise}.   Our proposal allows one to measure the average fidelity over all possible coherent states, using only  two-mode squeezing,  a 50-50 beamsplitter, and homodyne detection.
The same approach can be applied to benchmarks for  quantum  attenuation \cite{namiki2008fidelity,mivcuda2012noiseless,gagatsos2014heralded,zhao2017quantum,brewster2017noiseless}
and cloning \cite{cerf2000cloning,lindblad2000cloning,muller2012probabilistic} of coherent states, as well as the purification of displaced thermal states \cite{marek2007probabilistic,andersen2005experimental,zhao2017quantum}.



{\em General benchmark framework.} 
The scenario of quantum benchmarks  can be conveniently viewed as a game between an experimenter and a verifier \cite{yang2014certifying}.  The experimenter builds a device performing a quantum task, such as teleportation or cloning.  The verifier sets up a test in order to determine whether the device offers a quantum advantage.   The test consists in sending inputs to the device and performing measurements on the outputs.

Let us  start from the case of  a {\em deterministic device}, which  generates an output whenever it receives an  input.   Such a device can be described by a quantum channel (completely positive trace-preserving linear map),  transforming states of the input system into states of the output system.
Let us denote by $A$  ($A'$) the input (output) system, and by   $\map C$ a generic channel with input $A$ and output $A'$. 

\begin{figure}
\centering
\includegraphics[width=8cm]{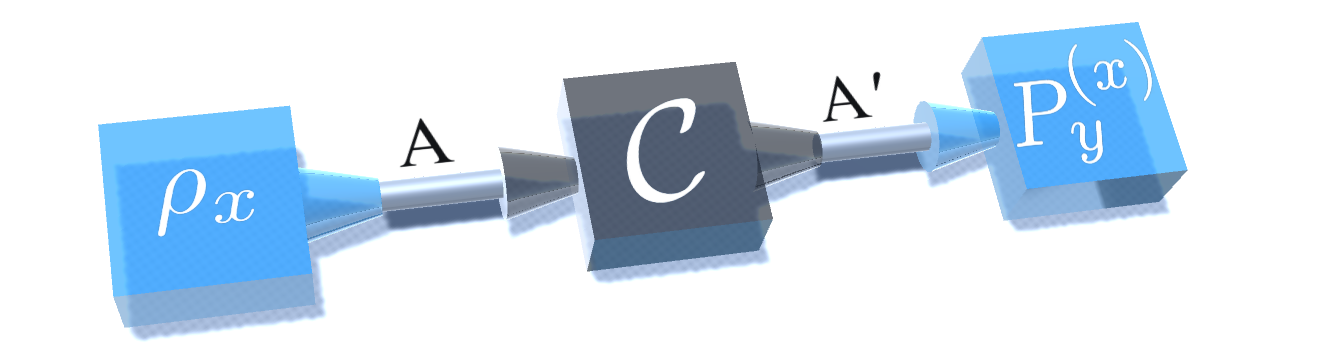}
\caption{{\bf Input-output test of a quantum device.} To test the device $\map C$, the verifier prepares an input state, randomly drawn from the set $\{\rho_x\}$.   Upon receiving the input, the device generates an output, which is then measured by the verifier with the POVM $\{  P^{(x)}_y\}$.  The  outcome is assigned  a score and the average score is used as a measure of performance. } \label{fig:general_benchmark_rhox}
\end{figure}

In order to rate the performance of the channel $\map C$, the verifier could use the setup described in Figure \ref{fig:general_benchmark_rhox}.  First, the verifier  prepares system $A$ in an input state $\rho_x$, randomly drawn from some set  $\{\rho_x\}$ with probability $p_x$. Then, the verifier  submits the input to the experimenter, who returns the output  $\map C(\rho_x)$. Finally,  the verifier performs a measurement, described by a positive operator-valued measure (POVM)  $\{P_y^{(x)}\}$ where  $x$ labels the measurement setting and $y$ labels the measurement outcome. For every setting $x$,  the outcome $y$ is  assigned a score $\omega(x,y)$. The average score
\begin{align}\label{scorefirst}
S^{\rm (det)}   =   \sum_x\sum_y    \,  \omega (x,y)  \,     p_x\,    \Tr \left[  P^{(x)}_y  \, \map C ( \rho_x)\right]
\end{align}
is then used as a figure of merit.   The typical example of Eq. (\ref{scorefirst}) is that of the    {\em fidelity-based benchmarks}
\cite{braunstein2000criteria,hammerer2005quantum,furusawa1998unconditional,rosenfeld2007remote,
boschi1998experimental,horodecki1999general,adesso2008quantum,owari2008squeezing,
chiribella2014quantum,zavatta2011high,namiki2008fidelity,chiribella2013optimal,
calsamiglia2009phase},  where the goal is to transform an unknown input state $\rho_x$ into a pure  target state $|\phi_x\>$. Fidelity benchmarks are expressed in terms of  the average fidelity
\begin{align}\label{fidelitybenchmark}
F^{\rm(det)} := \sum_x p_x  \,  \<\phi_x|\map{C}(\rho_x) |\phi_x\> \, ,
\end{align}
which can be viewed as the special case of  Eq. (\ref{scorefirst}) where each POVM  $\{  P^{(x)}_y\}$ has an outcome $y_x$ associated to the projector   $P^{(x)}_{y_x}  = |\phi_x\>\<\phi_x|$ and the score  $\omega (x,y)$  is either 1 or 0, depending on whether or not $y$ is equal to $y_x$.

The benchmark for a genuine quantum implementation has the form $F^{\rm(det)}  >  F^{\rm(det)}_{\rm C}$, where   $F^{\rm(det)}_{\rm C}$ is the {\em classical fidelity threshold}, namely the maximum fidelity achievable by measure-and-prepare channels \cite{hammerer2005quantum}.  The direct way to evaluate the score (\ref{scorefirst})---or the  average fidelity (\ref{fidelitybenchmark})---is to test the action of the channel  $\map C$ on all the input  states $\{\rho_x\}$ and to use the experimental data to compute the average. However,  this approach is not viable when the set of input states is infinite.  Now, we show that many  indirect ways to experimentally measure the average score (\ref{scorefirst}) or the  average fidelity (\ref{fidelitybenchmark}) exist. Among these indirect measurements, some can be dramatically simpler than the direct approach of Figure \ref{fig:general_benchmark_rhox}.

First of all,  we note that every test with random input states can be reformulated as a test with a single, {\em mixed}, input state $\sigma_{AR}$. This is because one can  regard the preparation of the state $\rho_x$ with probability $p_x$ as the preparation of  a single quantum-classical state  $\sigma = \sum_x p_x\rho_x \otimes \ketbra{x}_R$,
where  $R$ is an auxiliary system  keeping track of the index $x$.  Likewise, one can formally  write down a single quantum observable  $O  =  \sum_{x,y}  \,\omega (x,y)  \,   P_y^{(x)} \otimes \ketbra{x}_R $, so that the average score  (\ref{scorefirst}) takes the form
\begin{align}
S^{\rm(det)} := \Tr[O (\map C \otimes \map I_R)(\sigma_{AR})] \, .\label{eq:observable}
\end{align}

{\em Per se}, this reformulation does not make the problem easier.  The merit of Eq. (\ref{eq:observable}) is that it reveals a general structure, suggesting new ways to measure  the average score.    This reformulation  also offers a unified approach, which can be adopted not only for fidelity-based benchmarks, but also for other  types of quantum benchmarks, such as the   entanglement-based benchmarks  \cite{haseler2008testing,haseler2009probing,haseler2010quantum,killoran2012quantum}.

 \begin{figure}
\centering
\includegraphics[width=8cm]{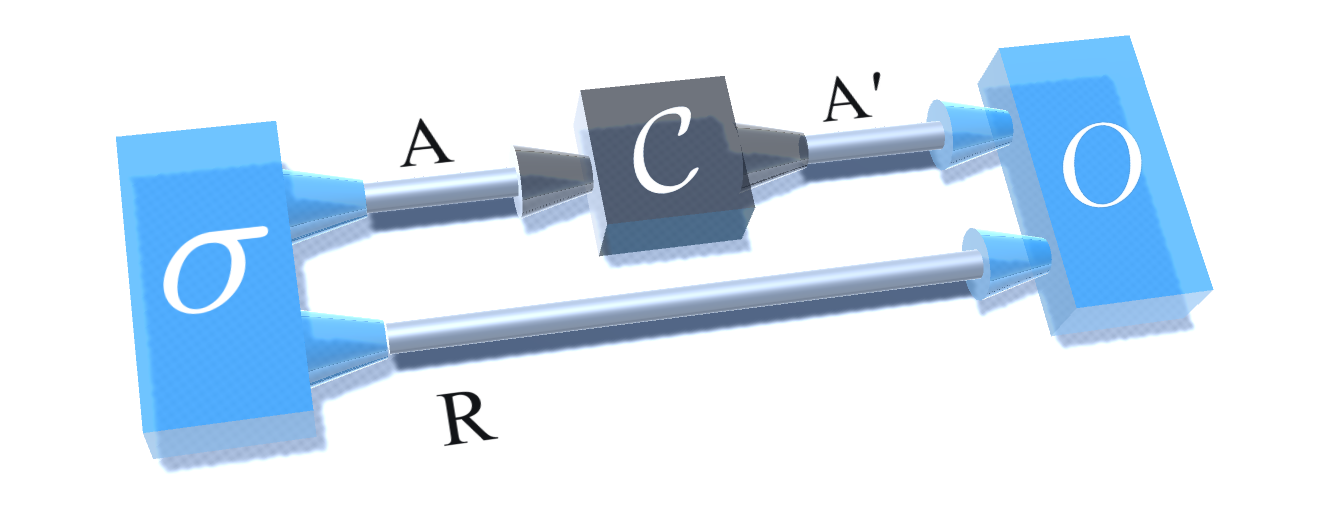}
\caption{{\bf Test with a single input and a single observable.} A composite system $AR$ in a joint state $\sigma$. Then, system  $A$ is sent to the device $\map C$, which transforms it into the output  system $A'$.  Finally, systems  $A'$ and $R$ undergo a joint measurement, described by the observable $O$.  The expectation value of $O$ is then used as the figure of merit. } \label{fig:general_benchmark}
\end{figure}

 The  single-input setup for testing quantum channels is depicted in Figure \ref{fig:general_benchmark}.   Now, the key observation is that many different tests are {\em equivalent}, meaning that they assign the same average score to all  possible channels.   This observation is important because, among the many equivalent tests, one can choose  the easiest to realize experimentally. Now, we develop a framework that captures the equivalence of tests and facilitates the search for the most convenient realization.
 The framework is based on the {\em \Jami{} operator} \cite{jamiolkowski1972linear}, defined as
\begin{align}
C := &~ \sum_{ij} \map{C}(\ket{i}\bra{j})
 \otimes  \ket{j}\bra{i} \end{align}
where $\{\ket{i}\}$ is a fixed  orthonormal basis for system $A$. In terms of the \Jami{}
operator, the average score can be written as 
 (Appendix \ref{app:COmega})
\begin{align}
S^{\rm(det)} = \Tr\left[ \Omega \,  C\right] \label{eq:simple_sdet}
\end{align}
where $\Omega$ is the operator on $A'A$ defined by
\begin{align}\label{performanceop}
\Omega
:= &~ \Tr_R\Big[\left(O_{A'R}  \otimes  I_A  \right) \, \left(I_{A'} \otimes \sigma_{AR} \right)  \Big] \, .
\end{align}
Here it is  understood that  the Hilbert spaces are rearranged in the appropriate order, so that the operators in the right hand side can be multiplied.

We call $\Omega$  the {\em performance operator} of the test.
For fidelity-based benchmarks, the performance operator is  simply the {\em average input-output state}
\begin{align}\label{perfid}
\Omega  =  \sum_x \,  p_x  \,  |\phi_x\>\<\phi_x|   \otimes \rho_x  \, .
\end{align}
where $\rho_x$ is the input and $\ket{\phi_x}$ is the target output.

{\em Canonical tests for deterministic devices.}  Clearly, two tests with the same  performance operator   are equivalent, even if they correspond to  totally different testing  procedures. Now, we exploit the equivalence to realize   every test through the preparation of a single pure state and the measurement of a single observable.

\begin{theo}
[Appendix \ref{app:equivbenchmark}]
\label{thm:equivbenchmark}
Every test for deterministic devices is equivalent to a canonical  test  of the following  form:
\begin{enumerate}
\item  Choose a mixed state $\tau_A$, with the property that the operator $I_{A'}\otimes \tau_A $ is invertible on the support of the   operator $\Omega^{T_A}$, where $T_A$ denotes the partial transpose on system $A$.
\item  Prepare a purification of $\tau_A$, denoted by $|\Psi\>_{AR}$.
\item Apply the channel  $\map C$ on system $A$.
\item Measure systems $A'$ and $R$  with  the observable
\begin{align}
O = &~ \left(I_{A'} \otimes \tau_R^{-1/2} \, T_{AR}^\dag  \right)\,   \Omega^{T_A} \,  \left(I_{A'} \otimes  T_{AR} \, \tau_R^{-1/2} \right) \, . \label{eq:equivobservable}
\end{align}
where $\tau_R = \Tr_A[\ketbra{\Psi}_{AR}]$ is the marginal of the state $\ket\Psi_{AR}$ on system $R$, and $T_{AR}$ is the partial isometry such that  $T_{AR}^\dag  \tau_A  T_{AR} = \tau_R$.
\end{enumerate}
\end{theo}
The best way to understand Theorem  \ref{thm:equivbenchmark} is to use  it in a concrete example.  Consider the problem of amplifying coherent states \cite{namiki2008fidelity,chiribella2013optimal,pooser2009low}.  Here, the task is  to transform a generic coherent state $\ket \alpha   \propto  \sum_n \,  \alpha^n  \,  |n\>/\sqrt{n!}$ into the amplified coherent state  $\ket{g\alpha}$, where  $g\geq 1$ is the gain of the amplifier.  For $g=1$, the problem is  to teleport coherent states  \cite{furusawa1998unconditional,sherson2006quantum,krauter2013deterministic} or  to store them in a quantum memory \cite{julsgaard2004experimental,lobino2009memory,hedges2010efficient,hosseini2011unconditional}.   Assuming that the inputs are Gaussian-distributed, the average fidelity is
\begin{align}\label{amplifid}
F^{\rm(det)}  =   \int  \frac{\d^2 \alpha}{\pi}  \, \lambda  e^{-\lambda|\alpha|^2} \,   \<g\alpha|  \,\map C (  |\alpha\>\<\alpha|)  \, |g\alpha\>  \,  ,
\end{align}
where $\lambda \geq 0$ is the inverse of the variance.
In practice, the average  cannot be evaluated directly, because this would require sampling over an infinite set of input states.  Moreover, in the actual experiments \cite{zavatta2011high}  the fidelity is evaluated through a full  tomography of the output state, meaning that each value of $\alpha$  requires a  large (ideally infinite) number of experimental settings, making the evaluation of the average fidelity prohibitively expensive.
  Luckily, Theorem \ref{thm:equivbenchmark} offers a  way out. Instead of sampling over all coherent states, it is enough to prepare a  two-mode squeezed vacuum state 
\begin{align}
\ket{\Psi}_{AR} = \sqrt{1-x} \sum_n x^{n/2} \ket{n}_A \otimes \ket{n}_R \label{eq:choosePsi} \, ,
\end{align}
where the squeezing parameter $x$ can be any number in the interval $(0,1)$. Instead of evaluating the fidelity on each coherent state, it  is enough to measure a single observable, given by Eq.  (\ref{perfid}) with the performance operator
\begin{align}\label{perfalpha}
\Omega  =  \int \frac{\d^2 \alpha}{\pi}  \,  \lambda e^{-\lambda|\alpha|^2} \,  |g\alpha\>\<g\alpha| \otimes |\alpha\>\<\alpha|  \, .
\end{align}

\begin{figure}
\centering
\includegraphics[width=9.5cm]{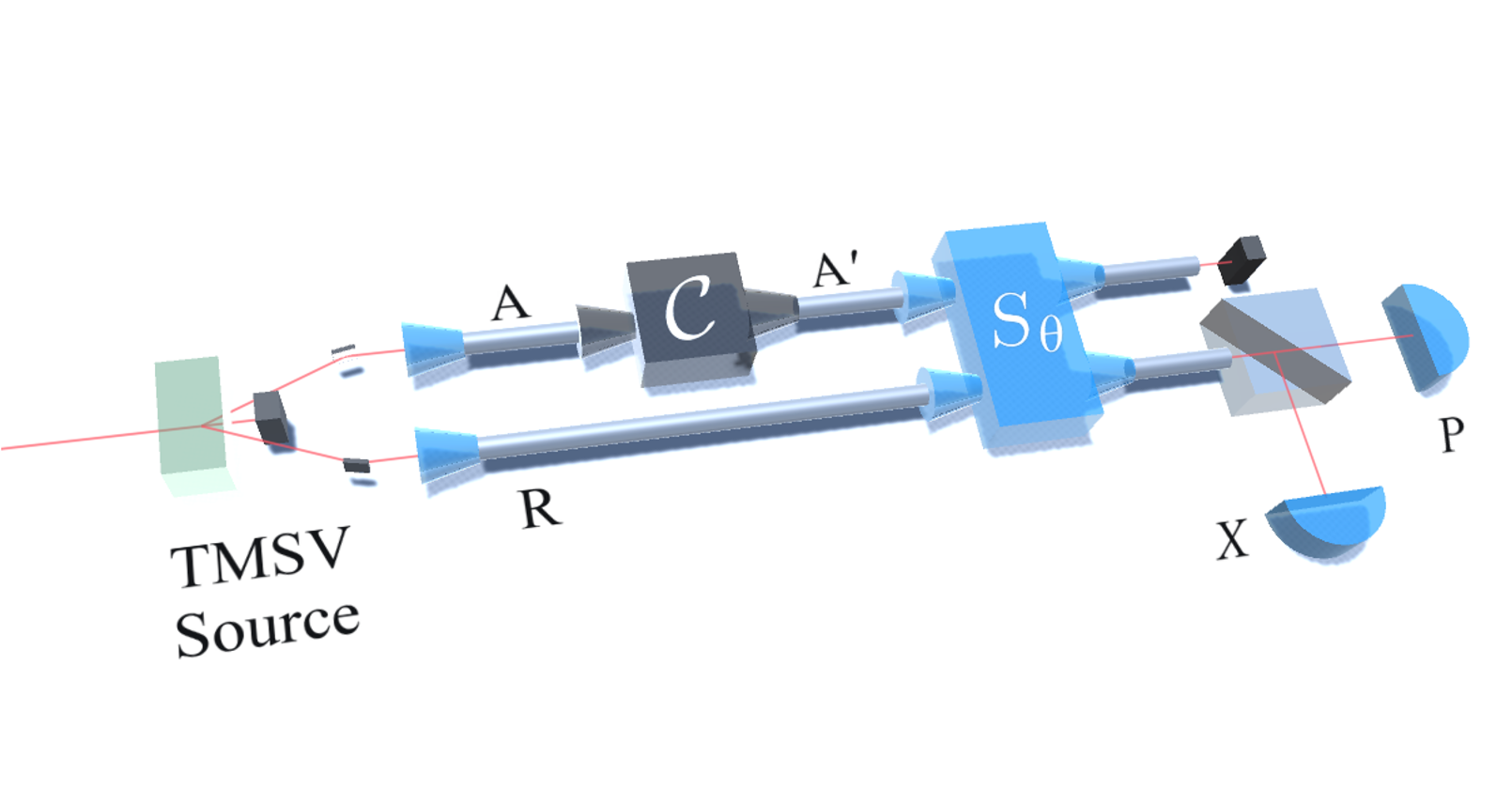}
\caption{{\bf Canonical test for coherent state amplifiers.}
The input mode and a reference are prepared in the two-mode squeezed vacuum (TMSV). After the action of the amplifier, the output mode and the reference  are sent through a two-mode squeezer $S_\theta$,  followed by  a 50-50 beamsplitter and two quadrature measurements on the output modes. }
\label{fig:amplifier_benchmark}
\end{figure}

Now, we take advantage of the fact that every value of the squeezing parameter $x$ is allowed, and therefore one can choose the most convenient  $x$.  Specifically, we notice  that the observable (\ref{eq:equivobservable}) takes a  simple form  when  $x = 1/(1+\lambda)$. For $g^2\le \lambda+1$,
 we find   
 (Appendix \ref{app:simplify1})
\begin{align}
O = &~ S^\dag_\theta  \,  (I\otimes G_\theta )  S_\theta \, ,  \label{eq:ampobs}
\end{align}
where  $S_\theta  =  \exp  [\theta  ( a b - a^\dag b^\dag )]$ is a two-mode squeezer with $\tanh \theta  =  g/ \sqrt{\lambda+1}$, and $G_\theta$ is the Gaussian observable $G_\theta=\sum_n  \,  (\tanh \theta)^{2n}\,  |n\>\<n|$.  In practice, this means that    the observable $O$ can be measured  by sending the two output modes  $A'$ and $R$ through a two-mode squeezer and by measuring the observable $G_\theta$ on the second port.  In turn, the observable $G_\theta$ can be  measured by sending the mode through  a 50-50 beamsplitter,  measuring the quadratures $X  =  (a+a^\dag)/2$ and $P  =  (b-b^\dag)/(2i)$ on the two output modes, respectively, and, finally, averaging the outcomes with a Gaussian weight (see 
Appendix \ref{app:simplify1} for the exact expression).
 The setup for $g^2 > \lambda+1$ is identical, except that one has to set $\tanh \theta =  \sqrt{\lambda+1}/g$ and the observable $G_\theta$ is measured on the first output port (Appendix \ref{app:simplify1}).

Our method makes the average fidelity  (\ref{amplifid})  experimentally accessible, thus enabling a rigorous experimental test of the  quantum advantage.   The same method can be  used to test the fidelity  of   attenuation \cite{namiki2008fidelity,mivcuda2012noiseless,gagatsos2014heralded,brewster2017noiseless},
cloning \cite{cerf2000cloning,lindblad2000cloning,muller2012probabilistic}, purification of displaced thermal states \cite{marek2007probabilistic,andersen2005experimental,zhao2017quantum},  and phase conjugation \cite{cerf2001phase},  as shown in 
Appendix \ref{app:simplify1} and \ref{app:simplify2}.
A limitation of the present approach  is that the verifier should be able to preserve the reference mode from noise.  In the case of quantum memories, this means that the verifier should possess a good quantum memory for the reference mode.  Basically, the test of Fig. \ref{fig:amplifier_benchmark} compares the untrusted quantum memory implemented by the experimenter with a  trusted quantum memory in the verifier's lab.

{\em Canonical tests for nondeterministic devices.}  Now, let us consider the case of devices that return an output with some nonunit probability. Examples of such devices are the noiseless probabilistic amplifier \cite{ralph2009nondeterministic}, experimentally realized in Refs.  \cite{kocsis2013heralded,zavatta2011high,ferreyrol2010implementation,usuga2010noise}, and the noiseless probabilistic attenuator  of Refs. \cite{mivcuda2012noiseless,gagatsos2014heralded,brewster2017noiseless}.
In general, a probabilistic device can be described by a quantum operation $\map C$ (completely positive trace-nonincreasing linear map). To test the device, one can  prepare a single input state $\sigma$ and measure an observable $O$ on the output, as in Figure \ref{fig:general_benchmark}. Sometimes, the device will report failure instead of producing an output.  The probability that an output is produced is
\begin{align}\label{psucc}
p_{\rm succ} = \Tr \Big[ (\map C\otimes \map I_R) (\sigma_{AR}) \Big]  =  \Tr  [ \map C (\sigma_A)] \, ,
\end{align}
where $\sigma_A = \Tr_R[\sigma_{AR}]$ is the marginal of $\sigma_{AR}$ on system $A$.   The average score is then
\begin{align}
 S^{\rm(prob)}  := \frac{\Tr[O(\map C \otimes \map I_R)(\sigma_{AR})]}{\Tr[\map C(\sigma_A)]}
\end{align}
and can be expressed as
\begin{align}\label{probscore}
S^{\rm(prob)} = \frac{\Tr[C\Omega]}{\Tr[C( I_{A'}\otimes \sigma_A)]}  \, ,
\end{align}
where $\Omega $ is the performance operator (\ref{performanceop}) and $C$ is the \Jami{} operator.  Note that, now, the score depends both on the performance operator $\Omega$ and on the marginal input state $\sigma_A$, which determines the probability of success via Eq. (\ref{psucc}).

It is easy to see that two tests are equivalent in terms of score {\em and} success probability if and only if they have the same pair of operators $(\Omega,\sigma_A)$.  Leveraging on the  equivalence, we can  construct a canonical realization.
\begin{theo}\label{theo:prob}
Every test of probabilistic devices is equivalent to a canonical test of the following form:
\begin{enumerate}
\item Prepare a purification of the marginal input state $\sigma_A$, denoted by $|\Phi\>_{A R}$.
\item Apply the quantum operation $\map C$ on system $A$.
\item Measure  systems $A'$ and $R$ with the observable
\begin{align}
O = &~ (I_{A'} \otimes \widetilde {\sigma}_R^{-1/2} \, T_{AR}^\dag) \, \Omega^{T_A} \, (I_{A'} \otimes   T_{AR}\, \widetilde {\sigma}_R^{-1/2}). \label{eq:equivobservableprob}
\end{align}
where $\widetilde {\sigma}_R$ is the marginal of the state $|\Phi\>_{AR}$ on system $R$ and  $T_{AR}$ is the partial isometry such that   $T_{AR}^\dag   \sigma_A  T_{AR}   =  \widetilde \sigma_R$.

\end{enumerate}
\end{theo}

Theorem \ref{theo:prob} offers the first rigorous way of testing the fidelity  benchmark for noiseless nondeterministic amplifiers \cite{kocsis2013heralded,zavatta2011high,ferreyrol2010implementation,usuga2010noise}.  In this case, the marginal state $\sigma_A$ is
\begin{align}
\sigma_A = \int \frac{\d^2\alpha}{\pi} \lambda e^{-\lambda|\alpha|^2} \,  |\alpha\>\<\alpha|  . \label{eq:barrhoamp}
\end{align}
Its purification is a  two-mode squeezed vacuum, given by Eq. (\ref{eq:choosePsi}) with $x=1/(1+\lambda)$.  Then, one can obtain the observable $O$ from Eqs. (\ref{eq:equivobservableprob}) and (\ref{perfalpha}). Again, the observable has a simple experimental realization. In fact, this is the same realization described in the deterministic case.  Using this realization,  it is now possible to set up a conclusive demonstration of quantum advantage for noiseless amplifiers.  The same holds for nondeterministic attenuation  \cite{mivcuda2012noiseless,gagatsos2014heralded,brewster2017noiseless}.




{\em The fully black box test.}  We analyzed, separately, the  tests of deterministic devices and the  tests of probabilistic devices. In practice, however, we  may not know   the success probability of the tested device.
This would be a problem, because the benchmark generally depends on the success probability  \cite{yang2014certifying}: in general,  the smaller the success probability, the higher the benchmark.   A solution to the problem would be to use the highest benchmark, calculated in the limit of vanishing success probability.   However, this could  set an unreasonably high bar for the experiment. Now, we show that the verifier can devise  a fully black box test, where the value of the benchmark is independent of the probability of success.

\begin{theo}
[Appendix \ref{app:fully_blackbox}]
 \label{thm:fully_blackbox}
Given a test   $\cal  T$ for deterministic devices, one can construct a new  test $\map T'$ for probabilistic devices, with the following properties:
\begin{enumerate}
    \item $\map T'$ has the same performance operator as the original test $\map T$. Therefore,  $\map T'$ assigns the same score as $\map T$ to all deterministic devices.
    \item For probabilistic devices, the benchmark for  $\map T'$ is independent of the  success probability.
\end{enumerate}
\end{theo}

The new test $\map T'$ is  described by a pair of operators $(\Omega, \sigma_A)$, with the following properties:  the performance operator $\Omega$ is chosen to be the same as the performance operator of the old test $\map T$.   This choice guarantees that the test $\map T'$ assigns the same score as $\map T$ when applied to deterministic devices.   The marginal state  $\sigma_A$ is chosen to be the state that reduces the probabilistic benchmark  to its minimum: this means that $\sigma_A$ minimizes the best score (\ref{probscore}) over  all measure-and-prepare channels.
The test for amplification or attenuation shown earlier in the article is an example of a fully black box test: the same experimental test and the same benchmark value can be used for both deterministic and probabilistic devices.   More examples of this situation are shown in
Appendix \ref{app:sym}, which  focusses on the scenario  where the test $\map T$ enjoys a symmetry with respect to a group of physical transformations. 



{\em Conclusions.}  In this article we showed that  a verifier can experimentally evaluate the performance of a quantum device on an infinite set of inputs, by preparing a single entangled input  and measuring a single joint   observable.   As an application, we constructed a test for  the  realization of quantum memories, amplifiers, and attenuators of coherent states, and purifiers  of displaced thermal states.    The test can be realized using  two-mode squeezers,  beamsplitters, and homodyne detection. Using these ingredients, one can experimentally assess the average fidelity  over all possible coherent states (or all possible displaced thermal states), thus  providing a fully rigorous  demonstration of  genuine quantum advantage.

\medskip

{\bf Acknowledgments.}   This work is supported by the National Natural Science
Foundation of China through Grant No. 11675136, by the Canadian Institute for Advanced Research
(CIFAR), by the Hong Kong Research Grant Council
through Grant No. 17326616,
by the HKU Seed Funding for Basic Research, and by the Foundational
Questions Institute through Grant No. FQXi-RFP3-1325.
\bibliography{benchmarks}

\appendix
\newcounter{appendixcounter}
\renewcommand{\theappendixcounter}{\Alph{appendixcounter}}
\newcommand{\appendixsection}[1]{\refstepcounter{appendixcounter}\section{Appendix \theappendixcounter: #1}}

\renewcommand{\theequation}{\theappendixcounter-\arabic{equation}}
\makeatletter
\@addtoreset{equation}{appendixcounter}
\makeatother

\appendixsection{Derivation of \Eq{eq:simple_sdet}} \label{app:COmega}

Here we show that the average score of the channel $\map C$ can be expressed in terms of the \Jami{} operator as  $S^{\rm(det)}  =  \Tr [C\, \Omega]$.

We remind the reader that the action of the channel can be expressed  in terms of the \Jami{} operator as follows:
\begin{align}
\nonumber \map{C}(\rho)   &=   \sum_{i,j}    \,  \rho_{ij}  \,      \map C ( |i\>\<j|)   \\
\nonumber    &  = \sum_{i,j}  \,  \map C (|i\>\<j|)  ~    \Tr\big[ \,  \rho \, |j\>\< i| \, \big]   \\
\nonumber &  =   \sum_{i,j}   \Tr_{A}  \Big[    \map C (|i\>\<j|)  \otimes \Big( \rho \,  |j\>\<i| \Big)   \Big] \\
&  =     \Tr_A\Big[ (  I_{A'} \otimes \rho ) \,  C \Big] \, .
\end{align}
The next step is to compute the average score. To keep track of the Hilbert spaces, we will add a subscript to each operator, so that, e.g.  $\rho_A$  indicates that the operator $\rho$ acts on the Hilbert space of system $A$.    We will also write expressions like    $(C_{A'A} \otimes I_R)  \, (O_{A'R} \otimes I_A) $,  with the implicit understanding that the Hilbert spaces are suitably reordered in order to perform the matrix multiplication.
With this notation, we have
\begin{align}
S^{\rm(det)} = & \Tr[O_{A'R}  \,   (\map C \otimes \map I_R)(\sigma_{AR})] \nonumber\\
 = & \Tr[O_{AR'}  \, \Tr_A[( I_{A'}\otimes \sigma_{AR})(C_{A'A} \otimes I_R)]] \nonumber\\
 = & \Tr[(O_{A'R} \otimes I_A) \,   (I_{A'}\otimes \sigma_{AR})  \,  (C_{A'A} \otimes I_R)] \nonumber\\
 = & \Tr[  (C_{A'A} \otimes I_R)  \, (O_{A'R} \otimes I_A) \,   (I_{A'}\otimes \sigma_{AR})  ] \nonumber\\
 = & \Tr[C_{AA'} \Tr_R[(O_{A'R} \otimes I_A) (  I_{A'} \otimes  \sigma_{AR)}] \nonumber\\
= & \Tr[C_{AA'} \, \Omega_{AA'} ] \label{eq:simple_sdet2}
\end{align}
with
\begin{align}\Omega_{AA'} : = \Tr_R[(O_{A'R} \otimes I_A) (  I_{A'} \otimes  \sigma_{AR)}]   \label{PerformanceOp}\, .
\end{align}

\appendixsection{Proof of Theorem \ref{thm:equivbenchmark}} \label{app:equivbenchmark}

Theorem \ref{thm:equivbenchmark} states that a (deterministic) test with performance operator $\Omega$ can be implemented through the preparation of a single pure state  and the measurement of a single joint observable.

To construct the pure state and the observable, we pick a state $\tau_A$ of system $A$, and we diagonalize it as
\begin{align}
\tau_A = \sum_n \lambda_n \ketbra {\phi_n}_A  \, .
\end{align}
We require that the operator $(I_{A'} \otimes \tau_A)$ is invertible on the support of the operator $O_{A'A}$, in such a way that operators like $(I_{A'} \otimes \tau)^{-1}  O_{A'A}$ are well defined.

The input state in our canonical test will be a purification of the state $\tau$, with purifying system $R$.    The purification, denoted by $|\Psi\>_{AR}$, can   be written in the Schmidt decomposition as
\begin{align}
\ket\Psi_{AR} =  \sum_n \sqrt{\lambda_n} \ket{\phi_n}_A \otimes \ket{\psi_n}_R \, ,
\end{align}
where the states $\{  |\psi_n\>\}$ are orthonormal.

The joint observable in the canonical test is defined as
\begin{align}\label{OAR!}
O_{A'R} = &~ (I_{A'} \otimes \tau_R^{-1/2}  \, T_{AR}^\dag  )\, \Omega_{A'A}^{T_A} \,   (I_{A'} \otimes  T_{AR} \, \tau_R^{-1/2}) \, ,
\end{align}

where
\begin{align}
\tau_R = \sum_n {\lambda_n} \ketbra{\psi_n}_R
\end{align}
 is the marginal of $\ket\Psi_{AR}$ on system $R$, and
 \begin{align}
 T_{AR}  =  \sum_n  |\phi_n\>_A  \<\psi_n|_R
 \end{align}
 is the partial isometry that maps each eigenvector of $\tau_R$ into the corresponding eigenvector of $\tau_A$.

Now, it remains to prove that the performance operator of our test is exactly $\Omega$.

Let us provisionally  denote  the performance operator of our test by $\Omega'$.   The goal is to show the equality  $\Omega' =  \Omega$.      For this purpose, the operator $\Omega'$ can be computed using Eq. (\ref{PerformanceOp}), with $\sigma_{AR}  =  |\Psi\>\<\Psi|_{AR}$ and $O_{A'R}$ defined in Eq. (\ref{OAR!}).   Explicitly, we have
\begin{align}
\Omega'_{A'A} = &~ \Tr_R\left[(O_{A'R} \otimes I_A) (  I_{A'} \otimes  |\Psi\>\<\Psi|_{AR}  ) \right] \nonumber\\
= &~ \Tr_{\widetilde A}\left[(\Omega^{T_{\widetilde A}}_{A'\widetilde A} \otimes I_A) (  I_{A'} \otimes  |\Gamma\>\<\Gamma|_{A\widetilde A}  ) \right]  \label{questa}\, .
\end{align}
where $\widetilde A$ is a second copy of system $A$, and $|\Gamma\>_{A\widetilde A}$ is the (unnormalized) vector  defined as
\begin{align}
\nonumber |\Gamma\>_{A\widetilde A}    &:  =     \left(I_A \otimes  T_{AR}   \tau_R^{-1/2} \right)  |\Psi\>_{AR}  \\
&  =  \sum_{n}     \,  |\phi_n\>_A  \otimes |\phi_n\>_{\widetilde A}
\end{align}

Continuing from Eq. (\ref{questa}), we obtain
\begin{widetext}
\begin{align}
\nonumber \Omega'_{A'A}    & = \sum_{m,n}  \,  \Tr_{\widetilde A}     \Big[\, \Big (\Omega^{T_{\widetilde A}}_{A'\widetilde A} \otimes I_A\Big)\  \Big (  I_{A'} \otimes |\phi_m\>\<\phi_n|_{\widetilde A} \otimes   |\phi_m\>\<\phi_n|_A  \Big)\, \Big] \\
\nonumber  &  =    \sum_{m,n} \,  \Big(  I_{A'} \otimes |\phi_m\>_A\<\phi_n|_{\widetilde A}   \Big)  \, \Omega^{T_{\widetilde A}}_{A'\widetilde A}  \,    \Big(  I_{A'} \otimes |\phi_m\>_{\widetilde A} \<\phi_n|_{A}  \Big)\\
\nonumber  &  =    \sum_{m,n} \,  \Big(  I_{A'} \otimes |\phi_m\>_A\<\phi_m|_{\widetilde A}   \Big)  \, \Omega_{A'\widetilde A}  \,    \Big(  I_{A'} \otimes |\phi_n\>_{\widetilde A} \<\phi_n|_{A}  \Big)\\
 & =  \Omega_{A'A} \, .
\end{align}
\end{widetext}
This concludes the proof of Theorem \ref{thm:equivbenchmark}.

\appendixsection{Canonical test for (noisy) coherent states}\label{app:simplify1}

Here we work out the explicit expression of the canonical test for  storage, teleportation, amplification, attenuation, cloning, and purification of (noisy) coherent states.

All the above can be subsumed into   a single task. In this task, the experimenter is given a displaced thermal state
\begin{align}
\rho_{\alpha,\mu} = \int \frac{\d^2 \beta}{\pi} \, \mu \,  e^{-\mu|\beta|^2} \ketbra{\alpha+\beta}  \, ,
\end{align}
where $\mu$ is a known parameter specifying the amount of noise in the input, while $\alpha$ is an unknown parameter specifying the modulation of the signal.
   The experimenter's  goal is to transform the displaced thermal state $\rho_{\alpha,\mu}$ into the pure coherent state $\ket{g\alpha}$, where $g$ is the gain of the amplification (or the attenuation parameter, if $g <1$).

  For $\mu \to \infty$,  the goal is to transform the pure coherent state $|\alpha\>$ into the coherent state $|g\alpha\>$.  Depending on whether $g>1$,  $g=1$, or $g<1$, this task is  amplification, teleportation/storage, or attenuation.
For finite $\mu$, the task is (ideally) to transform a  displaced thermal state into a pure coherent state, while amplifying, preserving, or attenuating the signal.
Finally, since $N$ copies of the  state $\rho_{\alpha,\mu}$ can be reversibly converted into a single copy of the state $\rho_{\sqrt N \alpha , \mu}$, the above task encompasses  various tasks of  cloning and purification.

The figure of merit is the average fidelity
\begin{align}\label{FAVE}
F  =  \int \frac{\d^2\alpha}{\pi} \,   \lambda \,  e^{-\lambda |\alpha|^2}  \,  \< g \alpha|  \,  \map C  (\rho_{\alpha,\mu}) \,  |g\alpha\>  \, ,
\end{align}
where $\map C$ is the channel used by the experimenter, and $p_\lambda (\d^2\alpha)  =  \lambda  e^{-\lambda |\alpha|^2} \,  \d^2\alpha/\pi$  is the probability distribution of the signal.
The performance operator of the fidelity test is
\begin{align}
\Omega = \int  \frac{\d^2\alpha}{\pi} \,   \lambda \,  e^{-\lambda |\alpha|^2}  \,  \ketbra{g\alpha} \otimes  \rho_{\alpha,\mu} \, ,
\end{align}
 consistently with  \Eq{perfid} of the main text.

 We now use  Theorem \ref{thm:equivbenchmark} to construct a new test for the average fidelity (\ref{FAVE}).
First of all, we have to choose a state $\tau_A$ such that $I_A'\otimes \tau_A$ is invertible on the support of $\Omega^{T_A}$.    Here we choose a generic   thermal state, decomposed as
\begin{align}\tau_A  =  (1-x)  \,  \sum_{n=0}^\infty  x^n  \,  |n\>\<n|_A \qquad x  \in  (0,1)  \, ,
\end{align}
where $\{  |n\>\}$ is the Fock basis.  The canonical test of Theorem \ref{thm:equivbenchmark}   uses a purification of $\tau_A$.   Specifically, we choose the two-mode squeezed vacuum
 \begin{align}
\ket{\Psi_x}_{AR} = \sqrt{1-x} \sum_{n=0}^{\infty} x^{n/2} \ket{n}_A \otimes \ket{n}_R  \, .
\end{align}
also known as the twin-beam state \cite{ferraro2005gaussian}.  Note that the purifying system $R$ is another Bosonic mode, and therefore $\spc H_R  \simeq \spc H_A$. The isomorphism is implemented by the unitary operator $T_{AR}  =  \sum_n  \, |n\>_A\<n|_R$.

  In the following we will construct  the observable $O_{A'R}$, using   \Eq{eq:equivobservable} of the main text.
  It is convenient to separate two  cases, depending on whether  the input states are pure or mixed.
\subsection{Pure inputs}

For pure input states, the observable $O_{A'R}$ of   \Eq{eq:equivobservable} reads
\begin{align}
O_{A'R} = &~ (I_{A'} \otimes \tau_R^{-1/2}  \,T_{AR}^\dag) \, \Omega^{T_A}_{A'A} \,   (I_{A'} \otimes \, T_{AR} \tau_R^{-1/2}) \nonumber\\
=&~ \int \frac{\d^2\alpha}{\pi}\, \lambda\, e^{-\lambda|\alpha|^2}\,    \ketbra{g\alpha}  \otimes \left(\tau_R^{-1/2} |\overline \alpha\>\<\overline \alpha|  \tau_R^{-1/2}\right) \nonumber\\
\nonumber =&~  \int \frac{\d^2\alpha}{\pi}\,    \frac {\lambda}{1-x} \,  e^{-  \left(\lambda + 1-\frac 1x \right)|\alpha|^2}   \\
  &  \qquad \times~   \ketbra{g\alpha}\otimes  \ketbra { \frac{\overline \alpha}{\sqrt x}}   \,.
\end{align}
The observable takes a simple form when
\begin{align}
x =  \frac 1{\lambda+1} \,,
\end{align}
in which case we have
\begin{align}\label{noiseless}
O_{A'R}   &=    \int \frac{\d^2\gamma}{\pi}\,  \ketbra{\frac{g \, \gamma}{\sqrt{\lambda+ 1}}}\otimes  \ketbra { \overline \gamma}  \, .
\end{align}   A more explicit expression comes from   the relation
\begin{align}
\nonumber   &  S_\theta\,  \Big(D(\alpha) \otimes D(\beta)  \Big) \,  S_\theta^\dag  \\
    & =   D\Big(  \cosh \theta  \, \alpha    - \sinh \theta \, \overline \beta \Big) \otimes  D\Big(\cosh \theta\,   \beta   -  \sinh \theta\,  \overline \alpha \Big)\,   \end{align}
where $S_\theta$ is the two-mode squeezing operation
\begin{align}
 S_\theta: =  \exp  \Big[\theta  (  a b  -  a^\dag b^\dag  )\Big] \, .
\end{align}
For $g \le  \sqrt{\lambda+1}$, we choose  \begin{align}
\theta  =  \tanh^{-1}     \left(  \frac g {\sqrt {\lambda+1}}\right)  \, ,
\end{align}
obtaining
\begin{align}
\nonumber O_{A'R}  &=   \int \frac{\d^2\gamma}{\pi}\,  S^\dag_\theta \left[  I  \otimes D \left(  \frac {\overline \gamma}{\cosh \theta}  \right)\right]   \,  \\
\nonumber & ~  \times   \Big( S_\theta  \ketbra 0  \otimes \ketbra 0\, S^\dag_\theta\Big)    \,  \left[I  \otimes D \left(   \frac{ \overline \gamma }{\cosh \theta} \right)\right]^\dag S_\theta  \\
\label{right}
  &  =  S_\theta^\dag   \Big (  I\otimes G_\theta  \Big ) \,S_\theta \, ,
\end{align}
where $G_\theta$ is  the Gaussian observable

\begin{align}\label{GaussObs}
G_\theta  &  =   \sum_n  \,   (\tanh \theta)^{2n}  \,  |n\>\<n|\, .
\end{align}

Similarly, for $g   >  \sqrt {\lambda +1}$, we choose  \begin{align}
\theta  =  \tanh^{-1}     \left(  \frac{\sqrt {\lambda+1}} {g}\right)  \, ,
\end{align}
obtaining
\begin{align}
O_{A'R} =
  (\tanh\theta)^2  S^\dag_\theta   \Big (  G_\theta\otimes I  \Big ) \,S_\theta \, ,  \label{left}
\end{align}
where  $G_\theta$ is the Gaussian observable defined in Eq. (\ref{GaussObs}), now with $\tanh \theta =  \sqrt{\lambda+1}/g$.

Eqs.  (\ref{right}) and (\ref{left})  imply that we can measure the observable $O_{A'R}$ by
\begin{enumerate}
\item performing the two-mode squeezing operation $S_\theta$ on the modes $A'$ and $R$
\item discarding one of the two output modes (the first mode, if $g  \le \sqrt {\lambda+1}$,  or the second mode, if $g  > \sqrt {\lambda+1}$), and measuring the Gaussian observable $G_\theta$ on the other.
\end{enumerate}
In turn, the measurement of the Gaussian observable $G_\theta$ can be implemented in different ways.    When $\tanh \theta$ is small, the observable $G$ can be accurately approximated using a photon counter that distinguishes  Fock states with low photon number.  In general, the Gaussian observable $G$ can be measured with a heterodyne setup,  corresponding to the POVM
\begin{align} P(\d^2 \gamma)   =   \ketbra \gamma  \,  \frac {\d^2 \gamma} \pi\, \,.
\end{align}
Upon obtaining the outcome $\gamma$, one can average the outcomes with the Gaussian weight
\begin{align}
w(\gamma)  = (\tanh\theta)^{-2} \, e^{-  \frac{|\gamma|^2}{ (\sinh \theta )^2}}   \, .
\end{align}
Finally, the heterodyne measurement can be implemented with two homodyne detectors, using  the following procedure
\begin{enumerate}
\item mix the target mode with the vacuum in a 50-50 beamsplitter
\item measure the quadrature $X  =  (a+a^\dag)/2$ on the first output mode and the quadrature $P=  (b-b^\dag)/(2i)$ on the second output mode
\item if the outcomes of the two quadrature measurements are $x$ and $p$, declare  the outcome  $\gamma  =   \sqrt 2  (x +  i  p)$.
\end{enumerate}
By construction, the expectation value of the above measurement is equal to the average fidelity of Eq. (\ref{FAVE}).

\subsection{Mixed inputs}
When the input states are mixed, the observable $O_{A'R}$ is
\begin{align}
O_{A'R} = &~ (I_{A'} \otimes \tau_R^{-1/2}  \,T_{AR}^\dag) \, \Omega^{T_A}_{A'A} \,   (I_{A'} \otimes \, T_{AR} \tau_R^{-1/2}) \nonumber\\
=&~ \int \frac{\d^2\alpha}{\pi}\, \lambda\, e^{-\lambda|\alpha|^2}\,    \ketbra{g\alpha}  \otimes \left(\tau_R^{-1/2} \rho_{\overline \alpha,\mu} \tau_R^{-1/2}\right) \nonumber\\
=&~  \int \frac{\d^2\alpha}{\pi}\,   \int \frac{\d^2\beta}{\pi}\,  \, \lambda\,  e^{-\lambda|\alpha|^2} \,  \mu \,  e^{-\mu| \beta|^2}   \nonumber \\
\nonumber & \times  \ketbra{g\alpha}\otimes \left( \tau_R^{-1/2} \ketbra{\overline {\alpha+\beta}} \tau_R^{-1/2}\right)\\
=&~  \int \frac{\d^2\alpha}{\pi}\,   \int \frac{\d^2\beta}{\pi}\,  \lambda\,   e^{-\lambda|\alpha|^2}  \, \mu\, e^{-\mu|\beta|^2} \frac{e^{-\lambda' |\overline{\alpha+\beta}|^2}}{1-x}   \nonumber\\
\nonumber & \qquad \qquad \times   \ketbra{g\alpha} \otimes \ketbra{\frac{\overline{\alpha+\beta}}{\sqrt{x}}}  \\
=&~ \int \frac{\d^2\alpha}{\pi}\,   \int \frac{\d^2\gamma}{\pi}\,  \lambda\,    e^{-\lambda|\alpha|^2} \,  \mu\,  e^{-\mu|\sqrt{x}   \gamma -  \alpha|^2} \, \frac{x  \,  e^{-\lambda' x |\gamma|^2}}{1-x} \nonumber \\&  \qquad  \qquad\times    \ketbra{g\alpha} \otimes \ketbra{\overline \gamma} \nonumber\\
=&~ \int \frac{\d^2\alpha}{\pi}\,   \int \frac{\d^2\gamma}{\pi}\,      (\lambda + \mu)  \, e^{-(\lambda+\mu)\left| \alpha - k  \gamma \right|^2}  \,  e^{ -l  |\gamma|^2}\nonumber  \\
& \qquad \qquad \times \ketbra{g\alpha} \otimes \ketbra{\overline \gamma} \, , \end{align}
with
\begin{align}
\label{k} k  & = \frac{\mu\sqrt{x}}{\lambda + \mu} \\
l  &= \mu x  + x-1   -  \frac{\mu^2  x}{\lambda+  \mu}\, .  \label{l}
\end{align}
We observe that the expression can be simplified if we set $l=0$, corresponding to the choice
\begin{align}
x=   \frac {\lambda+  \mu}{ \lambda+  \mu  +  \lambda\mu} \, .
\end{align}

\begin{figure}
\centering
\includegraphics[width=8.5cm, trim={4cm 0 0 0}]{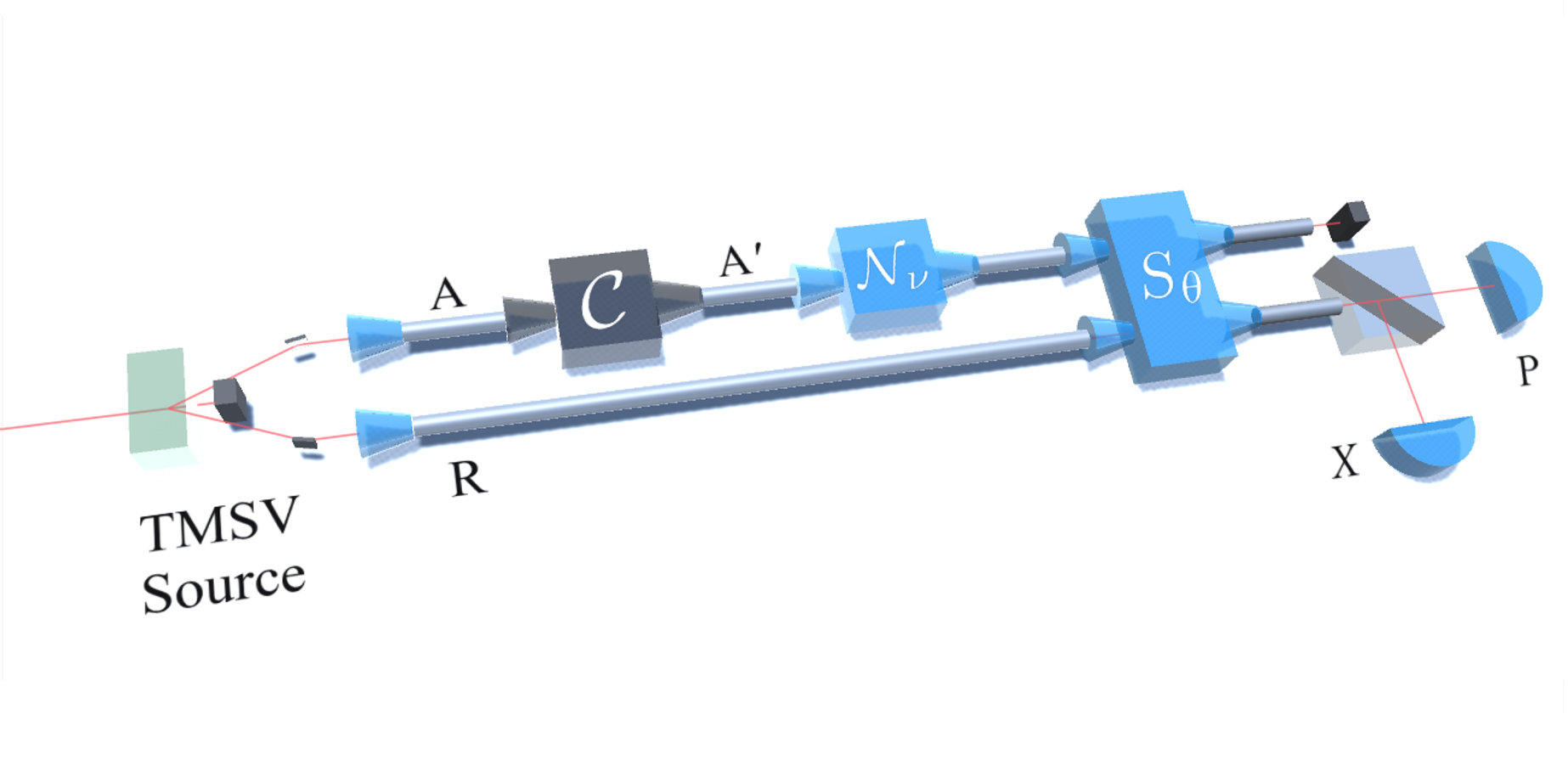}
\caption{{\bf Canonical test for the amplification/purification  of  noisy coherent states.}
The two input modes $A$ and $B$ are prepared in a two-mode squeezed vacuum state (TMSV), with suitably chosen squeezing parameter. Then,  system $A$ is input into the black box $\map C$.  Once the black box has acted,  the output mode $A'$ is  sent through the noisy channel $\map N_\nu$, the output of which is sent through a two-mode squeezer. Then, one mode is discarded and the other is sent through a 50-50 beamsplitter, after which the two quadratures $X$ and $P$ are measured. }
\label{fig:purifier_benchmark}
\end{figure}

Defining $\delta  =   g \, (  \alpha -  k \gamma)$, we obtain
\begin{align}
\nonumber O_{A'R} =&~   \int \frac{\d^2\gamma}{\pi}\,  \int \frac{\d^2\delta}{\pi}\,      (\lambda + \mu)  \, e^{-\frac{\lambda+\mu}{g^2}\left|\delta \right|^2}  \nonumber  \\
\nonumber & \qquad \qquad \times \ketbra{  \delta+  gk \gamma} \otimes \ketbra{\overline \gamma} \\
=&~     \int \frac{\d^2\gamma}{\pi}\,    \,    \map N (\ketbra{ gk  \gamma} )  \otimes \ketbra{\overline \gamma}  \, , \end{align}
where $\map N_\nu$ is the Gaussian-additive-noise  channel defined by
\begin{align}\label{noisyN}
\map N_\nu  (\rho)   =  \int \frac{\d^2\delta}{\pi}\,\nu  \, e^{-  \nu \, \left|\delta \right|^2}    \,     D  (\delta)  \,\rho \, D(\delta)^\dag
\end{align}
 and
 \begin{align}
 \nu  =    \frac{\lambda + \mu}{g^2}  \,.
 \end{align}
More concisely, the observable $O_{A'R}$ can be expressed as
\begin{align}
O_{A'R}   =       (\map N_\nu  \otimes \map I_R)    (   Z_{A'R}  ) \, ,
\end{align}
with
\begin{align}
Z_{A'R}   =   \int \frac{\d^2\gamma}{\pi}\,    \,  \,   \ketbra{ gk  \gamma}   \otimes \ketbra{\overline \gamma} \, .
\end{align}
Note that we have the relation
\begin{align}
\nonumber \Tr [O_{A'R}  \,  \rho  ]      & =  \Tr[   (\map N_\nu  \otimes \map I_R)    (   Z_{A'R}  )   \,  \rho ]  \\
& =  \Tr[  Z_{A'R}  \,  (\map N_\nu  \otimes \map I_R)    (   \rho ) ] \, ,
\end{align}
valid for every $\rho$.  Operationally, this means that  the measurement of  the observable $O_{A'R}$ can be realized by first applying the Gaussian channel $\map N_\nu$ and then measuring the observable $ Z_{A'R}$.  Also, note that the observable $Z_{A'R}$ has the same form of the observable $O_{A'R}$ in Eq. (\ref{noiseless}), with the only difference that  $1/\sqrt{\lambda+ 1}$ is now replaced by $k$.  Hence, we know that it can be  measured by performing a two-mode squeezing operation on modes $A'$ and $R$, discarding one of the modes, and measuring the single-mode Gaussian observable  $G_\theta$  [Eq. (\ref{GaussObs})] on the other mode.  Putting everything together, and using the homodyne realization of the observable $G_\theta$, we obtain the Gaussian setup  shown in Fig.  \ref{fig:purifier_benchmark}.

\appendixsection{Test for the complex conjugation  of (noisy) coherent states}
\label{app:simplify2}

In this section we design a test for the complex conjugation of coherent states, and for various combinations of this task with the tasks of storage, teleportation, amplification, attenuation, cloning, and purification.  As in the previous section, all the tasks in question can be subsumed into   a single task, where the experimenter has to transform a displaced thermal state  $\rho_{\alpha, \mu}$   into the pure coherent state $\ket{g\overline \alpha}$.

The figure of merit is the average fidelity
\begin{align}\label{FAVE2}
F  =  \int \frac{\d^2\alpha}{\pi} \,   \lambda \,  e^{-\lambda |\alpha|^2}  \,  \< g \overline\alpha|  \,  \map C  (\rho_{\alpha,\mu}) \,  |g\overline \alpha\>  \, ,
\end{align}
where $\map C$ is the channel used by the experimenter, and the performance operator of the fidelity test is
\begin{align}
\Omega = \int  \frac{\d^2\alpha}{\pi} \,   \lambda \,  e^{-\lambda |\alpha|^2}  \, \rho_{\alpha,\mu} \otimes    \ketbra{g\overline \alpha}\, .
\end{align}

For the input state, we choose the same two-mode squeezed state of the previous section.
Let us construct now the observable $O_{A'R}$, using  \Eq{eq:equivobservable} of the main text. Here, we have
\begin{align}
O_{A'R} = &~ (I_{A'} \otimes \tau_R^{-1/2}  \,T_{AR}^\dag) \, \Omega^{T_A}_{A'A} \,   (I_{A'} \otimes \, T_{AR} \tau_R^{-1/2}) \nonumber\\
=&~ \int \frac{\d^2\alpha}{\pi}\, \lambda\, e^{-\lambda|\alpha|^2}\,    \ketbra{g\alpha}  \otimes \left(\tau_R^{-1/2} \rho_{\alpha,\mu} \tau_R^{-1/2}\right) \nonumber\\
=&~  \int \frac{\d^2\alpha}{\pi}\,   \int \frac{\d^2\beta}{\pi}\,  \, \lambda\,  e^{-\lambda|\alpha|^2} \,  \mu \,  e^{-\mu|\beta|^2}   \nonumber \\
\nonumber
& \times  \ketbra{g\alpha}\otimes \left( \tau_R^{-1/2} \ketbra{\alpha+\beta} \tau_R^{-1/2}\right)  \\
=&~  \int \frac{\d^2\alpha}{\pi}\,   \int \frac{\d^2\beta}{\pi}\,  \lambda\,   e^{-\lambda|\alpha|^2}  \, \mu\, e^{-\mu|\beta|^2} \frac{e^{-\lambda' |\alpha+\beta|^2}}{1-x}   \nonumber\\
\nonumber & \qquad \qquad \times   \ketbra{g\alpha} \otimes \ketbra{\frac{\alpha+\beta}{\sqrt{x}}}  \\
=&~ \int \frac{\d^2\alpha}{\pi}\,   \int \frac{\d^2\gamma}{\pi}\,  \lambda\,    e^{-\lambda|\alpha|^2} \,  \mu\,  e^{-\mu|\sqrt{x}\gamma - \alpha|^2} \, \frac{x  \,  e^{-\lambda' x |\gamma|^2}}{1-x} \nonumber \\&  \qquad  \qquad\times    \ketbra{g\alpha} \otimes \ketbra{\gamma} \nonumber\\
=&~ \int \frac{\d^2\alpha}{\pi}\,   \int \frac{\d^2\gamma}{\pi}\,      (\lambda + \mu)  \, e^{-(\lambda+\mu)\left|\alpha - k\gamma \right|^2}  \,  e^{ -l  |\gamma|^2}\nonumber  \\
& \qquad \qquad \times \ketbra{g\alpha} \otimes \ketbra{\gamma} \, , \end{align}
with
\begin{align}
\label{k} k  & = \frac{\mu\sqrt{x}}{\lambda + \mu} \\
l  &= \mu x  + x-1   -  \frac{\mu^2  x}{\lambda+  \mu}\, .  \label{l}
\end{align}
Setting $l=0$  (corresponding to the choice $x=  (\lambda+  \mu)/(\lambda+ \mu + \lambda\mu)$) and changing variables, we obtain
\begin{align}
\nonumber O_{A'R} =&~ \int \frac{\d^2\delta}{\pi}\,   \int \frac{\d^2\gamma}{\pi}\,      (\lambda + \mu)  \, e^{-(\lambda+\mu)\left|\delta \right|^2}  \,   \nonumber  \\
\nonumber & \qquad \qquad \times \ketbra{g\delta+  gk\gamma} \otimes \ketbra{\gamma} \\
=&~   \int \frac{\d^2\gamma}{\pi}\,     \,   \map N_\nu (\ketbra{ gk\gamma} )  \otimes \ketbra{\gamma}  \, , \end{align}
where $\map N_\nu$ is the noisy channel defined by Eq. (\ref{noisyN}) and
\begin{align}
\nu  =  \frac {  \lambda+  \mu}{g^2} \, .
\end{align}

 \begin{figure}
\centering
\includegraphics[width=9cm, trim={3cm 0 0 0}]{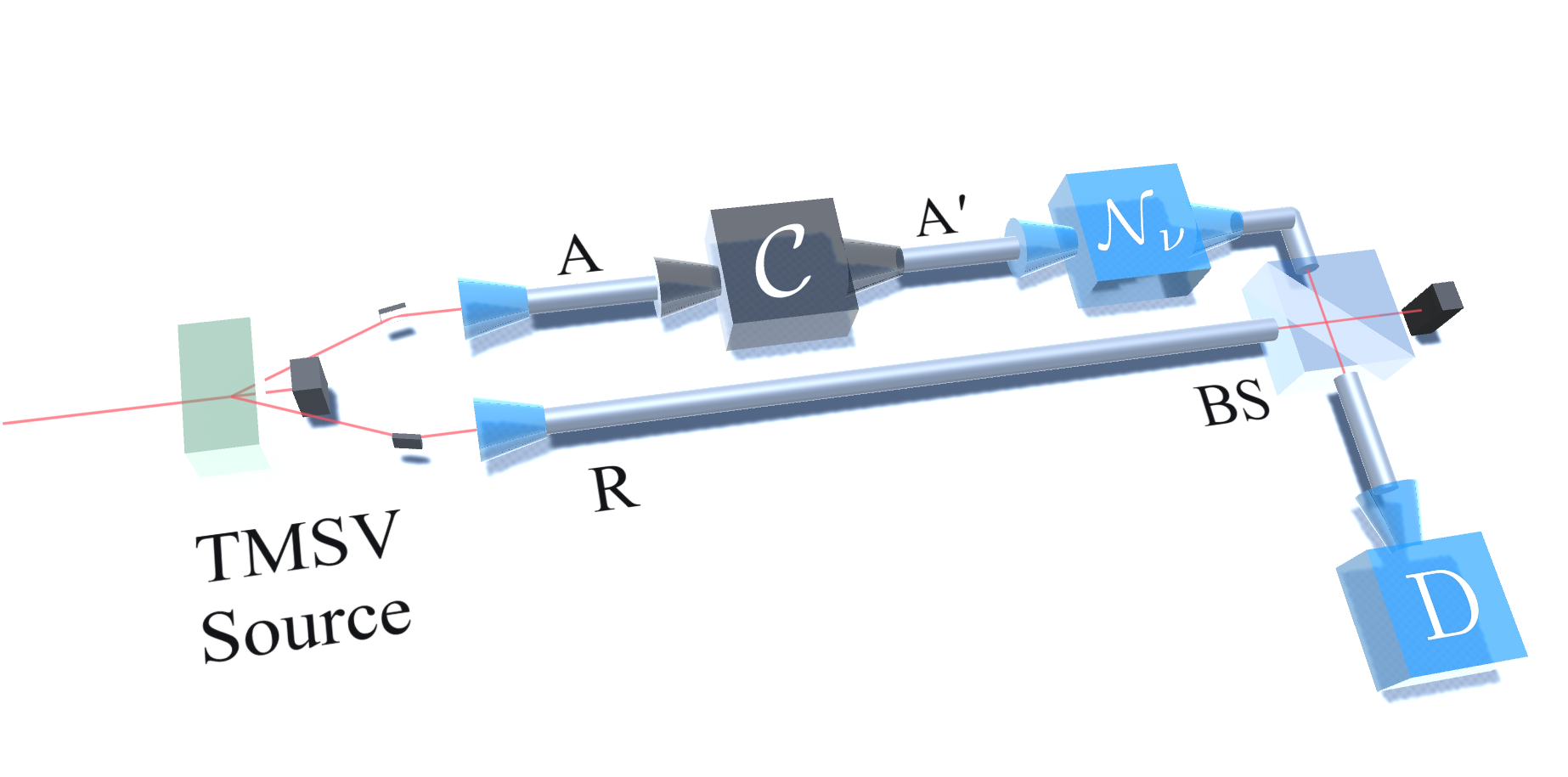}
\caption{{\bf Test for the complex conjugation of  noisy coherent states.}
The two input modes $A$ and $B$ are prepared in a two-mode squeezed vacuum state (TMSV).  Then,  system $A$ is input into the black box $\map C$.  Once the black box has acted,  the output mode $A'$ is  sent through the noisy channel $\map N_\nu$.  The output of the latter is sent through a beamsplitter together with mode $R$. Finally, mode $A'$ is discarded and mode $R$ is sent to a photodetector.  }
\label{fig:mixedconjugator_benchmark}
\end{figure}

More concisely, the observable $O_{A'R}$ can be expressed as
\begin{align}
O_{A'R}   =       (\map N_\nu  \otimes \map I_R)    (   Z_{A'R}  ) \, ,
\end{align}
with
\begin{align}
Z_{A'R}   =   \int \frac{\d^2\gamma}{\pi}\,    \,   \ketbra{ gk\gamma}   \otimes \ketbra{\gamma} \, .
\end{align}

In turn, the expression of the observable $ Z_{A'R}$  can be simplified using the following relation
\begin{align}
  &~ U_{gk}  \ket{gk\gamma} \otimes \ket{\gamma}  = \ket{\sqrt{g^2k^2+1}  \, \gamma}\otimes\ket0  \qquad \forall \gamma \in \C\, ,
\end{align}
where $U_{gk}$ is a suitable beamsplitter operator.
Using this relation, we obtain
\begin{align}
 \nonumber  & U_{gk}  \,   Z_{A'R} \,  U_{gk}^\dag \\
  =&~  \int \frac{\d^2\gamma}{\pi}\,    \,   \ketbra{  \sqrt{g^2k^2+1}  \,   \gamma}   \otimes   \ketbra{0}
   \nonumber\\
=&~   \frac 1 {g^2   k^2 +  1  }  \,   \Big (   I  \otimes |0\>\<0|  \Big)  \,.  \label{eq:i0int}
\end{align}

In summary, we constructed a procedure that allows us to experimentally measure the average fidelity (\ref{FAVE}) through the following steps
\begin{enumerate}
\item Prepare the two-mode squeezed  state  $|\Psi_x\>_{AR}$  with parameter $x=  (\lambda+ \mu)(\lambda+ \mu+ \lambda\mu)$.
\item  Apply the channel $\map C$ on the input mode $A$.
\item Apply the noisy channel $\map N$  [Eq. (\ref{noisyN})]  to  the output mode  $A'$.
\item  Let modes $A'$ and $R$ go through a beamsplitter described by the unitary operator $U_{gk}^\dag$, with $k$ as in Eq. (\ref{k}).
\item Discard mode $A'$ and send $R$ to photodetector.
\item
 If  no photon is detected, assign score  $1/(g^2  k^2  +1)$.  If one or more photons are detected, assign score 0.
\end{enumerate}
By construction, the expected frequency of the no detection events, divided by $g^2k^2+1$,  is equal to the average fidelity of Eq. (\ref{FAVE}).
  The procedure is illustrated in Figure \ref{fig:mixedconjugator_benchmark}.

\begin{figure}
\centering
\includegraphics[width=9cm, trim={3cm 0 0 0}]{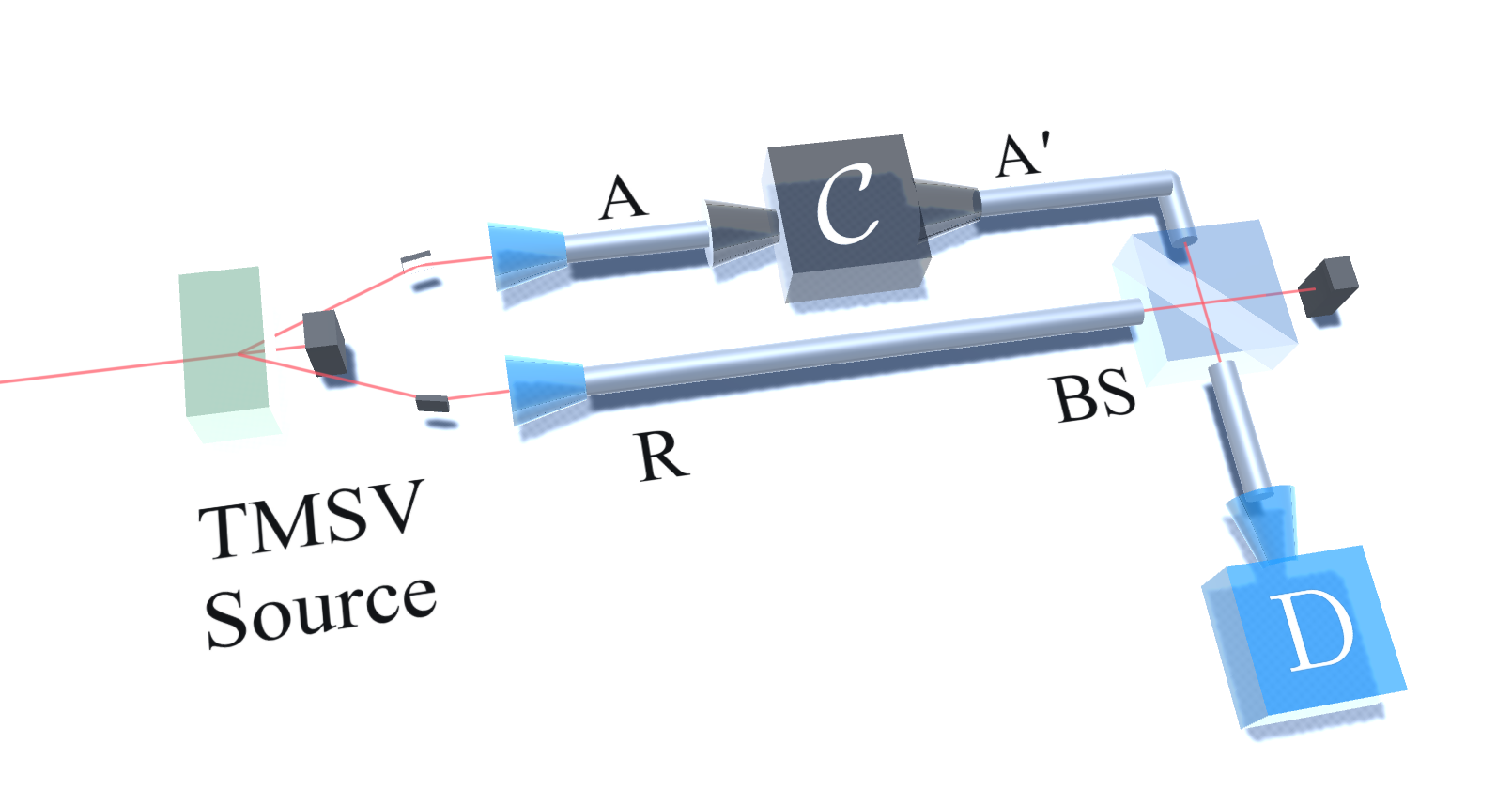}
\caption{{\bf Test for the complex conjugation of  pure coherent states.}
The two input modes $A$ and $B$ are prepared in a two-mode squeezed vacuum state (TMSV).  Then,  system $A$ is input into the black box $\map C$.  Once the black box has acted,  the output mode $A'$ is  sent  through a beamsplitter together with mode $R$. Finally, mode $A'$ is discarded and mode $R$ is sent to a photon counter.  }
\label{fig:pureconjugator_benchmark}
\end{figure}

Note that the third step (application of the channel $\map N_\nu$) is trivial  for pure input states. This is because the case of pure input  states corresponds to the limit $\mu \to \infty$, in which case Eq. (\ref{noisyN}) yields   $\nu \to \infty$ and $\map N_\nu  \to  \map I_A'$. The resulting setup is illustrated in Fig. \ref{fig:pureconjugator_benchmark}.

\appendixsection{Proof of Theorem \ref{thm:fully_blackbox}} \label{app:fully_blackbox}

To prove Theorem \ref{thm:fully_blackbox}, we have to show that, given a   test   $\cal  T$ for deterministic devices, one can construct a new  test $\map T'$ for probabilistic devices, with the following properties:
\begin{enumerate}
    \item $\map T'$ has the same performance operator as the original test $\map T$.
     \item The benchmark for  $\map T'$ is independent of the  success probability of the tested device.
\end{enumerate}

In the proof, we restrict our attention to tests represented by a performance operator $\Omega$ with positive partial transpose   (PPT), cf.  \cite{peres1996separability,horodecki2001separability}.  This can be done without loss of generality, as long as the performance operator $\Omega$ corresponds to a test where the experimenter prepares an input state $\sigma$ and measures a bounded observable  $O$.   In this case, one can always replace the original observable by the positive observable $O'  =  O  +  \|  O\|_\infty \,  I$, so that the resulting operator $\Omega'$  has PPT,  as one can  verify from the definition  (\ref{PerformanceOp}) using simple algebra.

For PPT performance operators, we have  two general expressions for the deterministic and probabilistic  benchmark:
\begin{lem} \label{lemma:values}
For a test with PPT performance operator $\Omega$, the maximum of $\Tr[C \,\Omega]$ over all  $C$'s that are \Jami{} operators of measure-and-prepare channels is
\begin{align}
  {S}^{\rm(det)}_{\rm M\& P}= &~ \inf_{\tau_A}  \Lambda^{\otimes} \Big [\left(I_{A'} \otimes \tau_A^{-1/2}\right) \, \Omega \,  \left(I_{A'} \otimes \tau_A^{-1/2} \right)   \Big]\, ,  \label{eq:optsdetmain}
\end{align}
where the infimum is taken over all states $\tau_A$ such that $I_{A'} \otimes \tau_A$ is invertible on the support of $\Omega$, and $\Lambda^{\otimes}$ is the {\em product numerical range} \cite{johnston2010family,gawron2010restricted}, defined as
\begin{align}
\Lambda^{\otimes} (O) = \sup_{\ket\alpha', \ket\alpha} \<\alpha'|\<\alpha|  O  |\alpha'\>  |\alpha\> \, ,
\end{align} the supremum being over all unit vectors $|\alpha'\>$ and $|\alpha\>$.
\end{lem}

\begin{lem}\label{lemma:values_prob}
For a test with PPT performance operator $\Omega$ and  marginal input state $\sigma_A$, the maximum of $\Tr  [ C\, \Omega ]/\Tr[C (I_{A'} \otimes \sigma_A)]$ over all $C$'s that are \Jami{} operators of measure-and-prepare quantum operations is
\begin{align}
  {S}^{\rm(prob)}_{\rm M\& P}= &~  \Lambda^{\otimes} \Big [\left(I_{A'} \otimes \sigma_A^{-1/2}\right) \, \Omega \,  \left(I_{A'} \otimes   \sigma_A^{-1/2} \right)   \Big]\,
 \label{eq:optsprobmain} \, .
\end{align}
\end{lem}
\medskip

The two lemmas are proven in the following subsections. Here we show how they  can be used to prove Theorem \ref{thm:fully_blackbox}.
\medskip

{\bf Proof of Theorem \ref{thm:fully_blackbox}.}  Let us see how to construct the test $\map T'$.   Let  $\tau_{\min}$ be the state that minimizes the right hand side of Eq. (\ref{eq:optsdetmain}).  Then, the pair $(\Omega, \tau_{\min})$ defines an equivalence class of  tests for probabilistic devices, with the equivalence relation defined in the main text (i.e. two tests are equivalent if they give the same score and the same probability of success for all quantum operations).  Now, consider the canonical test in this class, as defined by Theorem \ref{theo:prob} in the main text.    This is the desired test $\map T'$:  by construction, $\map T'$ has performance operator $\Omega$, which is the performance operator of $\map T$.  In addition,  Eqs. (\ref{eq:optsdetmain}) and (\ref{eq:optsprobmain}) imply that, for $\map T'$, the benchmark for arbitrary probabilistic devices (with arbitrarily small probability of success) coincides the benchmark for deterministic devices.  Hence, the benchmark for $\map T'$ is independent of the probability of success.   \qed

\medskip

In summary, the test $\map T'$ sets a single  threshold, independent of the success probability of the tested device. In the main text,  we called a test with this property a {\em fully black box test}.  Fully black box tests allow us to detect  quantum advantages  without knowing what is the probability  that the tested device produces an output.

The proof of Theorem \ref{thm:fully_blackbox} gives us a complete characterization of the fully black box tests:
\begin{cor}\label{cor:fully}
Let $\map T'$ be a test for probabilistic devices, with PPT performance operator $\Omega$ and marginal input state $\sigma_A$.    The test $\map T'$ is fully black box if and only if $\sigma_A$ is equal to $\tau_{\min}$, where $\tau_{\min}$ is the minimizer of the function
\begin{align}
f(\tau)  =   \Lambda^{\otimes} \Big [\left(I_{A'} \otimes \tau_A^{-1/2}\right) \, \Omega \,  \left(I_{A'} \otimes \tau_A^{-1/2} \right)   \Big] \, .
\end{align}
\end{cor}

In the following sections, we give the proof of Lemmas \ref{lemma:values} and \ref{lemma:values_prob}.

\subsection{Proof of Lemma \ref{lemma:values}}

We have to compute  the maximum of $\Tr [C \,\Omega ]$ over all  \Jami{} operators $C$ corresponding to measure-and-prepare  channels.   To this purpose, we use the technique developed in the proof  of Theorem 1 in \cite{chiribella2013optimal}. Since $\map C$ is a channel, the operator $C$ must satisfy the condition $\Tr_{A'}[C] = I_A$\cite{horodecki2003entanglement}.
 Since  in addition the channel $\map C$ is measure-and-prepare, the operator $C$ must be positive and separable \cite{horodecki2003entanglement}.
To handle the separability condition, we use the following property:
\begin{prop}\cite{doherty2004complete}\label{thm:hierarchy}
A positive operator $C$ on $\map H_{A'} \otimes \map H_A$ is separable if and only if for every $ n \in \mathbb{N}$ there exists  a positive operator $ C_n$ on $\map H_{A'}^{\otimes n} \otimes \map H_A$ s.t.
\begin{enumerate}
\item $C_n$ is an extension of $C$, namely $\Tr_{n-1}[C_n] = C$, where $\Tr_{n-1}$ denotes partial trace over first $n-1$ copies of $\map H_{A'}$;
\item $C_n$ is symmetric on $\map H_{A'}^{\otimes n}$, namely $(\Pi_n \otimes \map I_A)C_n = C_n$, where $\Pi_n$ is the permutation-twirling
    $$\Pi_n = \frac{1}{n!} \sum_{\pi \in S_n} U_\pi \rho U_\pi^\dagger$$
    where $U_\pi$ is a unitary operator that permutes the $n$ modes of $\map H_{A'}$ according to $\pi$.
\end{enumerate}
\end{prop}

Using the above result, the supremum over all  \Jami{} operators of measure-and-prepare channels can be written as
\begin{align}
&~{S}^{\rm(det)}_{\rm M\& P} \nonumber\\
= &~ \sup_{C} \Tr\left[C \,   \Omega \right] \nonumber\\
=&~ \inf_{n\in \mathbb N} \sup_{
{\footnotesize \begin{array}{ll}
&C_n:     C_n \geq 0,\\
 & (\Pi_n \otimes \map I_A)\, C_n = C_n \, ,\\
 & \Tr_n[C_n]=I_A
 \end{array}} }
\Tr\Big[\Tr_{n-1}[C_n]   \,  \Omega \Big] \nonumber\\
=&~ \inf_{n\in \mathbb N} \sup_{
{\footnotesize \begin{array}{ll}
&C_n:     C_n \geq 0,\\
 & (\Pi_n \otimes \map I_A)C_n = C_n,\\
 & \Tr_n[C_n]=I_A
 \end{array}} } \Tr\left[C_n \left(I_{A'}^{\otimes (n-1)} \otimes \Omega\right) \right] \nonumber\\
=&~ \inf_{n\in \mathbb N}   \sup_{
{\footnotesize \begin{array}{ll}
&C'_n:     C'_n \geq 0,\\
 & \Tr_n[C'_n]=I_A
 \end{array}} }
 \Tr\left[C'_n ~  (\Pi_n \otimes \map I_A) \left(I_{A'}^{\otimes (n-1)} \otimes \Omega\right) \right] , \label{eq:separableChoi}
\end{align}
where $C_n'$ is a generic (not necessarily permutationally invariant) operator.

Now, the optimization over $C'_n$ is  a  semidefinite program.    Its optimal value is equal to the optimal value of the dual program
\begin{align}
\inf &~ \Tr[\Lambda_n] \nonumber\\
\text{s.t.} &~ (I_{A'}^{\otimes n} \otimes \Lambda_n) \geq (\Pi_n \otimes \map I_A) (I_{n-1} \otimes \Omega)\, .
\end{align}
Note that, since $\Lambda_n$ is an arbitrary positive operator acting on system $A$, and since $\Omega$ has PPT, the dual program can be equivalently written as
 \begin{align}
\inf &~ \Tr[\Lambda_n] \nonumber\\
\text{s.t.} &~ (I_{A'}^{\otimes n} \otimes \Lambda_n) \geq (\Pi_n \otimes \map I_A) (I_{n-1} \otimes \Omega^{T_A})\, ,
\end{align}
with the advantage that now the operator $\Omega^{T_A}$ is positive.
Hence, we can express the quantum benchmark as
\begin{align}\label{sdp-dual}
{S}^{\rm(det)}_{\rm M\& P} = &~ \inf_{n\in \mathbb N} \inf_{\Lambda_n: (I_{A'}^{\otimes n} \otimes \Lambda_n) \geq (\Pi_n \otimes \map I_A) (I_{n-1} \otimes \Omega^{T_A})} \Tr[\Lambda_n].
\end{align}

We also observe that, since  $\Omega^{T_A}$ is positive, the condition
\begin{align}(I_{A'}^{\otimes n} \otimes \Lambda_n) \geq (\Pi_n \otimes \map I_A) (I_{n-1} \otimes \Omega^{T_A})\end{align}
guarantees that  the operator on the left hand side is invertible on the support of the operator on the right hand side.
Note also that the operator $\Lambda_n$ must be nonnegative and therefore it can be written as
\begin{align}
\Lambda_n   =   \lambda_n\,   \tau_n \, ,
\end{align}
where $\tau_n$ is a density operator on the Hilbert space $\spc H_A$ and $\lambda_n$ is a non-negative constant.

Under this fact,  we can rewrite the average score as
\begin{align}
&~{S}^{\rm(det)}_{\rm M\& P} \nonumber\\
= &~ \inf_{n\in \mathbb N} \inf_{\tau_n > 0, \Tr[\tau_n]=1} \nonumber \\&~ \inf\{\lambda_n: \lambda_n (I_{A'}^{\otimes n} \otimes \tau_n) \geq (\Pi_n \otimes \map  I_A) (I_{n-1} \otimes \Omega^{T_A})\} \nonumber\\
= &~\inf_{n\in \mathbb N} \inf_{\tau_n > 0, \Tr[\tau_n]=1} \nonumber \\&~ \inf\{\lambda_n: \lambda_n (I_{A'}^{\otimes n} \otimes I_A) \geq (\Pi_n \otimes \map I_A) (I_{n-1} \otimes \Omega^{T_A}_{\tau_n}) \} \label{infimum} \, ,
\end{align}
with
\begin{align}
\Omega^{T_A}_{\tau_n} = \left( I_{A'}^{\otimes n} \otimes \tau_n^{-1/2}\right)\,  \Omega^{T_A} \,\left(I_{A'}^{\otimes n} \otimes \tau_n^{-1/2}\right) \, .
\end{align}
Note that the infimum over $\lambda_n$ in Eq. (\ref{infimum}) is the operator norm of the operator  $(\Pi_n \otimes \map I_A) (I_{n-1} \otimes \Omega^{T_A}_{\tau_n})$.  Hence, we can rewrite the average score as
\begin{align}
&~{S}^{\rm(det)}_{\rm M\& P} \nonumber\\
= &~\inf_{n\in \mathbb N} \inf_{\tau_n > 0, \Tr[\tau_n]=1} \| (\Pi_n \otimes \map I_A) (I_{n-1} \otimes \Omega_{\tau_n})\|_\infty   \,.
\end{align}
Note also that $\tau_n$ is a generic density operator on the Hilbert space $\spc H_A$, and therefore the dependence on $n$ can be removed: one has
\begin{align}
&~{S}^{\rm(det)}_{\rm M\& P} \nonumber\\
= &~\inf_{n\in \mathbb N} \inf_{\tau > 0, \Tr[\tau]=1} \| (\Pi_n \otimes \map I_A) (I_{n-1} \otimes \Omega^{T_A}_{\tau})\|_\infty   \,,
\end{align}
where $\tau$ is a generic density operator on the Hilbert space $\spc H_A$.  Continuing the chain of equalities, we get
\begin{align}
&~{S}^{\rm(det)}_{\rm M\& P} \nonumber\\
= &~\inf_{n\in \mathbb N} \inf_{
\footnotesize
\begin{array}{c}\tau > 0\\
 \Tr[\tau]=1
 \end{array}} \| (\Pi_n \otimes \map I_A) (I_{n-1} \otimes \Omega^{T_A}_{\tau})\|_\infty \nonumber\\
= &~\inf_{
\footnotesize \begin{array}{l}\tau > 0\\
 \Tr[\tau]=1
 \end{array}}   \inf_{n\in \mathbb N}  \| (\Pi_n \otimes \map I_A) (I_{n-1} \otimes \Omega^{T_A}_{\tau})\|_\infty \nonumber\\
= &~\inf_{
\footnotesize
\begin{array}{c}\tau > 0\\
 \Tr[\tau]=1
 \end{array}}   \inf_{n\in \mathbb N} \sup_{
 \footnotesize \begin{array}{c}
 \rho_n \geq 0 \\
 \Tr[\rho_n] = 1
 \end{array}} \Tr[\rho_n (\Pi_n \otimes \map I_A) (I_{n-1} \otimes \Omega^{T_A}_{\tau})] \nonumber \\
= &~\inf_{\footnotesize
\begin{array}{c}\tau > 0\\
 \Tr[\tau]=1
 \end{array}}  \inf_{n\in \mathbb N} \sup_{\footnotesize
 \begin{array}{c}
 \rho_n \geq 0 \\
  \Tr[\rho_n] = 1
  \end{array}} \Tr[\, (\Pi_n \otimes \map I_A) (\rho_n)   ~(I_{n-1} \otimes \Omega^{T_A}_{\tau}) \,]  \nonumber \\
= &~\inf_{\footnotesize
\begin{array}{c}\tau > 0\\
 \Tr[\tau]=1
 \end{array}}  \inf_{n\in \mathbb N} \sup_{
 \footnotesize
 \begin{array}{c}
 \rho'_n \geq 0  \\
  (\Pi_n \otimes \map I_A) (\rho'_n) =  \rho_n'   \\
   \Tr[\rho'_n] = 1
   \end{array}} \Tr[  \rho_n'   \,(I_{n-1} \otimes \Omega^{T_A}_{\tau}) ] \nonumber \\
= &~\inf_{\footnotesize
\begin{array}{c}\tau > 0\\
 \Tr[\tau]=1
 \end{array}}  \inf_{n\in \mathbb N} \sup_{\footnotesize
 \begin{array}{c}
 \rho'_n \geq 0  \\
  (\Pi_n \otimes \map I_A) (\rho'_n) =  \rho_n'   \\
   \Tr[\rho'_n] = 1
   \end{array}} \Tr[  \Tr_{n-1} [\rho_n'] \,    \Omega^{T_A}_{\tau} ]\nonumber \\
= &~\inf_{\footnotesize
\begin{array}{c}\tau > 0\\
 \Tr[\tau]=1
 \end{array}}
 \inf_{n\in \mathbb N} \sup_{\rho  \in   \set{Ext} (n;A',A) } \Tr[  \rho \,    \Omega^{T_A}_{\tau}) ] \, ,
\end{align}
where $ \set{Ext} (n;A',A)$ is the set of all density operators on $\spc H_{A'}\otimes \spc H_A$ that admit a symmetric extension on the space   $\spc H_{A'}^{\otimes n} \otimes \spc H_A$.

Using Property \ref{thm:hierarchy}, we conclude that the average score is the maximum over all {\em separable} density operators, namely
\begin{align}
{S}^{\rm(det)}_{\rm M\& P}
&~ =\inf_{\tau > 0, \Tr[\tau]=1} \sup_{\footnotesize
\begin{array}{c}
\rho\geq 0  \\
 \Tr[\rho] = 1 \\
 \rho \text{ separable}
 \end{array}} \Tr[\rho \Omega^{T_A}_{\tau}] \nonumber\\
\nonumber &~ = \inf_{\tau > 0, \Tr[\tau]=1}  \Lambda^{\otimes} ( \Omega^{T_A}_{\tau} ) \\
&~ = \inf_{\tau > 0, \Tr[\tau]=1}  \Lambda^{\otimes} ( \Omega_{\tau} ) \, .
\end{align}

This  concludes the proof of Lemma \ref{lemma:values}.

\subsection{Proof of Lemma \ref{lemma:values_prob}}

For any probabilistic measure-and-prepare device $\map C$ with \Jami{} operator $C$, we define
\begin{align}
\nonumber \Gamma =&~ I_{A'} \otimes \sigma_A\\
\nonumber \rho =&~ \frac{\Gamma^{1/2} C \Gamma^{1/2}}{\Tr[\Gamma^{1/2} C \Gamma^{1/2}]} \\
\Omega_{\sigma_A}   = &~  \Gamma^{-1/2}  \Omega  \Gamma^{-1/2} \, .
\end{align}
Note that $\rho$ is a quantum state.   Then, we write the score of $\map C$ as
\begin{align}
{S}^{\rm(prob)} = & \frac{\Tr[C \Omega]}{\Tr[C\Gamma]} \nonumber\\
= & \Tr[\Gamma^{-1/2} \, \rho \,  \Gamma^{-1/2} \Omega] \nonumber\\
= & \Tr[\rho \,  \Gamma^{-1/2} \Omega \Gamma^{-1/2}] \nonumber\\
= & \Tr[\rho  \, \Omega_{\sigma_A}] \nonumber\\
\leq & \sup_{\rho\text{ separable}}\Tr[\rho\,  \Omega_{\sigma_A}] \nonumber\\
= & ~ \Lambda^{\otimes} (\Omega_{\sigma_A}) \label{eq:probperformance}
\end{align}
The last inequality holds because for classical $\map C$, $C$ is separable, and thus $\rho$ is separable, too. \\

Eq. (\ref{eq:probperformance}) gives an upper bound on the benchmark ${S}^{\rm(prob)}_{\rm M\& P}$. It remains to prove that, in fact, the upper bound holds with the equality sign. To this purpose,  pick the unit vectors $|\psi\>  \in \spc  H_{A'} $ and $|\phi\>_{A}\in \spc H_A$ such that \begin{align}
\sup_{\rho\text{ separable}}    \Lambda^{\otimes}  ( \Omega_{\sigma_A}) =  \bra\psi  \bra\phi   \Omega_{\sigma_A}  \ket\psi  \ket\phi  \, .
\end{align}
Let
\begin{align}
Q = &~ \frac{\sigma_A^{-1/2} \ketbra\phi   \sigma_A^{-1/2}}{\Tr\left[\sigma_A^{-1/2} \ketbra\phi \sigma_A^{-1/2}\right]},
\end{align}
and we pick the \Jami{} operator as
\begin{align}
C = &~  \ketbra\psi\otimes Q
\end{align}
where $C$ correspond to a probabilistic measure-and-prepare device that performs the projective measurement POVM $\{Q,I-Q\}$, and, if the outcome corresponds to the projector $Q$,  outputs the state  $\ketbra\psi$.

By construction, the probabilistic measure-and-prepare device with \Jami{} operator $C$ achieves score
\begin{align}
\nonumber
&\frac{\Tr[C\,  \Omega]}{\Tr [ C  \,  (I_A'  \otimes \sigma_A)]}   \\
\nonumber   &  = \frac{\Tr [  (  \ketbra{\psi}  \otimes \sigma_A^{-1/2} \ketbra\phi \sigma_A^{-1/2} )  \Omega]}{\Tr  [  (\ketbra{\psi}\otimes \sigma_A^{-1/2} \ketbra\phi \sigma_A^{-1/2})  \,  (I_{A'} \otimes \sigma_A)]}  \\
\nonumber
&   =\Tr  [    ( \ketbra{\psi}  \otimes   \ketbra\phi  )  \,  \Omega_{\sigma_A}] \\
  &= ~ \Lambda^{\otimes} (\Omega_{\sigma_A}) \, .
\end{align}
This concludes the proof of Lemma \ref{lemma:values_prob}.

\appendixsection{Tests with symmetry}\label{app:sym}

In this section we consider tests that exhibit a symmetry with respect to a group of physical transformations.  We also provide further examples of fully black box tests, where the value of the benchmark is independent of the probability of success of the tested setup.

\subsection{Definitions and examples}
In the following we will assume that a certain group  of physical transformations $\grp G$  acts on the systems $A$ and $A'$. For example, $A$ and $A'$ could be qubits and the group $\grp G$ could be the group of rotations of the Bloch sphere, or the group of rotations around the $z$ axis.
 More generally, $A$ and $A'$ could be two systems of different dimension.   We denote by $(   U_g)_{g \in  \grp G}$ and  $(   U'_g)_{g \in  \grp G}$ the two unitary (projective) representations of $\grp G$ acting on the Hilbert spaces $\spc H_A$ and $\spc H_{A'}$, respectively.

Using the above notation, we  define what it means for a test to have  symmetry:
\begin{definition}{\bf (Covariant  tests for deterministic devices)}
Let $\map T$ be a test for deterministic devices  and let $\Omega \in  \spc H_{A'} \otimes \spc H_A$ be the corresponding performance operator.
We say that the test $\map T$ is \emph{covariant} with respect to the action of $\grp G$ iff the performance operator $\Omega$ satisfies the condition
\begin{align}\label{eq:symmetry_cond}
[\Omega,   U'_g  \otimes U_g ]   =     0  \qquad \forall g\in\grp G \, .
\end{align}
\end{definition}

When the group $\grp G$ is compact, an example of covariant test is the fidelity test  for the transformation $\rho_g  \to     |\psi_g\>\<\psi_g|$, where $g$ is chosen at random according to the normalized Haar measure $\d g$ and the states are defined as
\begin{align}
\rho_g  :  =  U_g \, \rho  \, U_g^\dag  \, , \qquad {\rm and} \quad  |\psi_g\>  := U_g'  \,  |\psi\>  \, ,
\end{align} $\rho$ being a fixed density matrix and $|\psi\>$ being a fixed unit vector.  In this case, the performance operator  is
\begin{align}
\Omega   =  \int \,  \d g\,   \ketbra{\psi_g}  \otimes \rho_g
\end{align}
and satisfies Eq. (\ref{eq:symmetry_cond}) due to the invariance of the Haar measure.

Two examples of tests with symmetry are presented in the following:
\begin{enumerate}
 \item {\bf Fidelity test  for the teleportation of pure states.}   Consider the task of transmitting a generic pure state $|\psi\>$ of a $d$-dimensional system. The fidelity test for the teleportation of pure states \cite{boschi1998experimental,yang2014certifying} has performance operator
\begin{align}
\nonumber \Omega   & =  \int \d \psi  \,     \ketbra{\psi} \otimes \ketbra{\psi}  \\
  &  =   \frac{P_{+}}{\Tr[P_+]} \, ,
\end{align}
where $P_+$ is the projector on the symmetric subspace,   and $\d \psi$ is the normalized invariant measure on the pure states.    In this case, the performance operator satisfies the condition $[\Omega ,  U\otimes U]  =  0$ for arbitrary unitary gates.

\item {\bf CHSH test.} It is important to stress that a test can be covariant even if its is not testing  a transformation of the form $\rho_g  \to     |\psi_g\>\<\psi_g|$.  As an example, consider the following entanglement-based test,  designed to test the preservation of a Bell inequality:
\begin{enumerate}
\item prepare an input qubit and a reference qubit in the entangled state $|\Phi^+\>  =  (  |0\>\otimes |0\>  +   |1\>  \otimes |1\>  )/\sqrt 2$
\item send the input qubit through the tested device
\item test the CHSH inequality on the output qubit and the reference.
\end{enumerate}
The CHSH test corresponds to the two-qubit observable
\begin{align}
\nonumber O    &=   Z  \otimes  \frac{  Z+  X}{\sqrt 2}   +   Z  \otimes   \frac{  Z  -X}{\sqrt 2}   \\
\nonumber & \quad  +  X  \otimes  \frac{  Z+  X}{\sqrt 2}   -  X  \otimes  \frac{  Z-X}{\sqrt 2}  \\
&  =\sqrt 2    \left (\frac{  Z+X}{\sqrt 2}    \otimes  \frac{    Z+  X}{\sqrt 2}     + \frac{  Z-X}{\sqrt 2}    \otimes  \frac{   Z-X}{\sqrt 2}\right)  \, .
\end{align}
 Using Eq.  (\ref{PerformanceOp}), one obtains the performance operator
 \begin{align}\label{CHSH}
 \Omega  =   \frac O 2  \, ,
 \end{align}
 where the operator $O$ on the right hand side is interpreted as acting on qubits $A'$ and $A$.    It is easy to see that the performance operator satisfies the commutation relations
 \begin{align*}
\qquad~  [ \Omega, X\otimes X  ]=  0 \, , \quad [ \Omega, Y\otimes Y  ]=  0 \, , \quad  [ \Omega, Z\otimes Z  ]=  0 \, ,
 \end{align*}
 meaning that the test is covariant with respect to the action of the Pauli group.
  \end{enumerate}

The definition of covariant tests, formulated in the deterministic case, can be extended  to the probabilistic case:
\begin{definition}{\bf (Covariant tests for probabilistic devices)}
Let $\map T$ be a test for probabilistic devices, let $\Omega$ be the  performance operator, and let $\sigma_A$ be the marginal input state on system $A$.
We say that the test $\map T$ is \emph{covariant} with respect to the action of $\grp G$ iff the performance operator $\Omega$ satisfies the condition \Eq{eq:symmetry_cond} and the marginal state $\sigma_A$ satisfies the condition
\begin{align}
[\sigma_A , U_g]  =  0\, . \label{eq:probinv}
\end{align}
\end{definition}

For a compact group $\grp G$, a fidelity test for the transformation $\rho_g \to \ketbra{\psi_g}$ (with $g$ chosen according to the Haar measure)   is covariant. This is because the marginal input state is
\begin{align}
\sigma_A   =  \int \d g  \,     \rho_g \, ,
\end{align}
which obviously satisfies the relation    $[\sigma_A , U_g]  =  0$.
   In the teleportation example, the marginal input state is
\begin{align}
\sigma_A  =  \frac {I} d  \, ,
\end{align}
where $I$ is the identity matrix on $\spc H_A$ and $d$ is the dimension of $\spc H_A$.  Clearly, the state $\sigma_A$  commutes with every unitary operator.

Again, it is important to stress that there are other examples of covariant tests other than the tests for   transformations of the form $\rho_g  \to     |\psi_g\>\<\psi_g|$.  For example, the CHSH test, viewed as a test on probabilistic devices, is also covariant (with respect to the Pauli group):  indeed, the marginal input state is
\begin{align}
\sigma_A   =    \Tr_R   [  |\Phi^+\>\<\Phi^+|]  =  \frac  I  2  \, ,
\end{align}
 and clearly commutes with the Pauli matrices $X, Y$, and $Z$.

\subsection{Examples of fully black box tests}

Here we show a class of examples where it is easy to construct the fully black  box test.  In all these examples, the original test is covariant and the representation acting on the input system is irreducible.     We recall that the representation $(U_g)_{g\in\grp G}$ is called irreducible if no subspace is invariant under its action, except for the trivial subspace $\{0\}$ and the whole Hilbert space $\spc H_A$.    The Schur's lemma then guarantees that, for every operator $X$, the condition
\begin{align}
[  X,  U_g]   =  0  \qquad \forall g\in\grp G
\end{align}
implies that $X$ has the form
\begin{align}
X  =  c\,  I_A  \, ,
\end{align}
where $c \in  \C  $ is a suitable constant.
\begin{lem}\label{lem:irr}
Let $\map T$ be a test for deterministic devices with PPT performance operator $\Omega$. Suppose that the input system $A$ has  dimension $d<\infty$.   If $\map T$  is covariant under the action of $\grp G$ and if and the group representation acting  on  $A$ is irreducible, then the deterministic benchmark is
\begin{align}\label{bench-irr}
S^{\rm(det)}_{\rm M\&P}   =   d\,    \Lambda^{\otimes} (\Omega) \, .
\end{align}
Every test $\map T'$ with  performance operator $\Omega$ and marginal input state $\sigma_A  =  I_A/d$ is a fully black box test.
\end{lem}

\Proof{}    To derive Eq. (\ref{bench-irr}) we use the dual program
\begin{align*}
{S}^{\rm(det)}_{\rm M\& P} = &~ \inf_{n\in \mathbb N} \inf_{\Lambda_n: (I_{A'}^{\otimes n} \otimes \Lambda_n) \geq (\Pi_n \otimes \map I_A) (I_{n-1} \otimes \Omega^{T_A})} \Tr[\Lambda_n] \, ,
\end{align*}
coming from  Eq. (\ref{sdp-dual}).     Observe that the right hand side of the inequality on $\Lambda_n$ commutes with the group representation $(U_g^{\prime \, \otimes n}  \otimes U_g)_{g\in\grp G}$.  By twirling both sides of the inequality with respect to this representation, we obtain a new operator $\Lambda'_n$, which commutes with each $U_g$ and has the same trace of $\Lambda_n$. Hence, the benchmark can be rewritten as
\begin{align*}
{S}^{\rm(det)}_{\rm M\& P} =  &~  \inf_{n\in \mathbb N} \inf_{\footnotesize
\begin{array}{c}
\Lambda'_n: (I_{A'}^{\otimes n} \otimes \Lambda'_n) \geq (\Pi_n \otimes \map I_A) (I_{n-1} \otimes \Omega^{T_A})  \\
{[}\Lambda'_n,  U_g{]}   =  0  \quad \forall g \in\grp G
 \end{array}}
\,  \Tr [\Lambda'_n] \, .
\end{align*}
Now, the Schur's lemma implies that $\Lambda_n'$ is proportional to the identity matrix. We write it as $\Lambda_n'   =   \lambda_n  \,    I/d$.    The rest follows by substituting $\tau$ with $I/d$ in the proof of Lemma \ref{lemma:values}: following the steps of the proof we obtain
\begin{align}
S^{\rm(det)}_{\rm M\& P} &~ = \inf_{\tau > 0, \Tr[\tau]=1}  \Lambda^{\otimes} ( \Omega^{T_A}_{\tau} ) \, ,
\end{align}
 with $\Omega^{T_A}_\tau  =  (I_{A'} \otimes \tau^{-1/2}) \, \Omega^{T_A} \, (I_{A'} \otimes \tau^{-1/2})$.   Substituting $\tau =  I/d$ we then obtain the desired expression
\begin{align}
\nonumber S^{\rm(det)}_{\rm M\& P} & = d \,   \Lambda^{\otimes} ( \Omega^{T_A} )  \\
&  = d \,   \Lambda^{\otimes} ( \Omega )  \, .
\end{align}
Now, note that, by construction, the infimum over $\tau$ is achieved by the maximally mixed state $\tau_{\min}=  I/d$.    Hence, the characterization of Corollary \ref{cor:fully} implies that the pair $(\Omega,  I/d)$ defines a fully black box test.  \qed
\medskip

It is useful to  illustrate the theorem in a few  examples.

\begin{enumerate}
\item {\bf Fidelity test for the teleportation of pure states.}   Consider  the fidelity test for the teleportation of arbitrary pure states in dimension $d$.   As we have seen in the previous Section, this test has operators
\begin{align}
\Omega  =   \frac{ P_+}{ \Tr[  P_+]}   \qquad {\rm and}  \qquad  \sigma_A  = \frac I d  \,
\end{align}
and is covariant under the action of the group   $U(d)$ of all unitary operators in dimension $d$.  Note that the representation of the group on the input system is irreducible.
Using Lemma \ref{lem:irr} we obtain the benchmark
\begin{align}
\nonumber S^{\rm(det)}_{\rm M\& P} &~ = d \,   \Lambda^{\otimes} ( \Omega )     \\
 \nonumber  & ~  =    d   \,   \sup_{\||\alpha\>\|  = 1 \, ,  \| |\beta\> \|=1} \,  \<\alpha|  \<\beta|  \Omega  |\alpha\>|\beta\>  \\
\nonumber   &  ~=  \frac {d}{\Tr[ P_+]}\\
  &  ~  =   \frac 2{d+1} \, .
\end{align}
This value coincides with the known fidelity benchmark for pure states  \cite{boschi1998experimental,yang2014certifying}.  Since the state $\sigma_A$ is maximally mixed, we know that the fidelity test is fully black box, meaning that the benchmark  $2/(d+1)$ holds independently of the probability of success of the tested device.

\item {\bf The CHSH test.}  Another interesting example of fully black box test is the entanglement-based CHSH test, also described in the previous Section.   The CHSH test has operators
\begin{align}
\nonumber \qquad \Omega~  &= \frac 1{\sqrt 2}    \left (\frac{  Z+X}{\sqrt 2}    \otimes  \frac{    Z+  X}{\sqrt 2}     + \frac{  Z-X}{\sqrt 2}    \otimes  \frac{   Z-X}{\sqrt 2}\right)  \\
\qquad ~\sigma_A   & =  \frac I 2 \, .
\end{align}
In this case, the test is  covariant under the action of the Pauli group, which is irreducible on the input space.
 The performance operator is not PPT, but can be transformed into a PPT operator by adding a constant term proportional to $ I\otimes I$. Since this transformation only offsets the product numerical range by a constant, we still can use  Lemma \ref{lem:irr} to compute the benchmark, obtaining
  \begin{align}
\nonumber ~\qquad S^{\rm(det)}_{\rm M\& P} &~ = 2 \,   \Lambda^{\otimes} ( \Omega )     \\
 \nonumber  & ~  =    2   \,   \sup_{\||\alpha\>\|  = 1 \, ,  \| |\beta\> \|=1} \,  \Tr \Big[\, \Big(\ketbra{\alpha} \otimes \ketbra{\beta} \Big)\,   \Omega \Big] \, ,
\end{align}
 which can be evaluated explicitly using the Bloch representation
 \begin{align}
 \nonumber
 \ketbra{\alpha}   &=  \frac {  I  +    m_x \,  \widetilde X  +  m_y\,  \widetilde Y  +    m_z\,  \widetilde Z }2  \\
 \ketbra{\beta}   &=  \frac {  I  +    n_x \,   \widetilde X  +  n_y\,  \widetilde Y  +    n_z\,  \widetilde Z }2  \, ,
    \end{align}
  $\st m  = (m_x,m_y,m_z)$ and $\st  n =  (  n_x, n_y  n_z)$ are unit vectors in $\R^3$, and $\widetilde X,  \widetilde Y, \widetilde Z$
are defined as follows:   \begin{align}
 \widetilde X  =  \frac{  Z+X}{\sqrt 2} \, , \quad \widetilde Z  = \frac{  Z-X}{\sqrt 2}\, , \quad \widetilde Y =  Y  \, .
 \end{align}
  Using this notation, the performance operator can be rewritten as
  \begin{align}
  \Omega  =  \frac { \widetilde X  \otimes  \widetilde X  +   \widetilde Z  \otimes  \widetilde Z}{\sqrt 2}
  \end{align}
  and one has
  \begin{align}
  \nonumber S^{\rm(det)}_{\rm M\& P} &~ =  \sqrt 2  ~  \Lambda^{\otimes}   (\widetilde X  \otimes  \widetilde X  +   \widetilde Z  \otimes  \widetilde Z  )  \\
 \nonumber  &~= \sqrt 2  ~   \sup_{\st m,  \st n}    \,  \{  m_x n_x  +  m_z  n_z\}  \\
  &  ~= \sqrt 2 \, .
  \end{align}
Note that the benchmark is strictly smaller than the Bell inequality value, which is equal to 2.   The reason is that here we are restricting the optimization over the set of two-qubit separable states, while the measurements  are fixed.    Lemma \ref{lem:irr} guarantees that preparing a maximally entangled input state, letting the unknown device act, and measuring the CHSH observable on the output is a fully black box test: any experimental value above $\sqrt 2$ guarantees that the tested  device has performance above the performance of every measure-and-prepare device, even allowing measure-and-prepare devices that postselect on some subset of favourable outcomes.

\item {\bf Fidelity test for the teleportation of pure states on the equator of the Bloch sphere.}

Consider the set of qubit pure states
\begin{align}\label{eq-states}
\ket{\phi_k} = \frac{1}{\sqrt2} \left( \ket0 + e^{2\pi i k / N} \ket1\right) \, ,
\end{align}
with $k  \in  \{0,1,\dots,  N-1\}$ and $ N  >2$  (we exclude the trivial case $N=2$, in which the states are orthogonal and the teleportation task can be achieved perfectly by measuring the input state).
The above states  are generated by the action of the cyclic group $C_N$ on the state $|\phi_0\>  = (|0\>  + |1\>)/\sqrt 2$. The group action
represented by the unitary matrices
\begin{align}
U_k := e^{  \frac{2\pi  i k }{N} \, Z}    \, .
\end{align}
Consider the fidelity test for the teleportation of the states (\ref{eq-states}).   In such test, the verifier prepares a state $|\phi_k\>$ chosen  with uniform probability $p_k  = 1/N$, lets the tested device act, and finally measures the fidelity with the state $|\phi_k\>$.

According to the general formula for fidelity tests,  the performance operator is
\begin{align}
\nonumber \Omega   &=  \frac 1 N  \,  \sum_{k}   \ketbra{\phi_k}  \otimes \ketbra{\phi_k}   \\
\nonumber  &  =  \frac  1{4 } \,   \Big(   \ketbra  0 \otimes \ketbra 0  +    \ketbra 1 \otimes \ketbra 1  \\
 \nonumber   &  \qquad  \qquad    +   2  |\Psi^+\>\<\Psi^+|  \Big) \, ,
\end{align}
with $|\Psi^+\>    =   (  |0\>  \otimes |1\>    +   |1\>  \otimes |0\> )/\sqrt 2$.   On the other hand, the marginal input state reads
\begin{align}
\nonumber \sigma_A   &=  \frac 1  N  \,  \sum_k  \, \ketbra{\phi_k}  \\
  &  =  \frac {I}2 \, .
\end{align}
Here there is an interesting point to make.    The test is based on a teleportation task, corresponding to a transformation of the form  $\rho_g \to \ketbra{\psi_g}$, where $g$ is an element of the cyclic group.  As a consequence (see the previous Section),  the test is covariant under the action of the cyclic group.   But in fact, the symmetries of the test are even larger:   Indeed, it is easy to check that one has
\begin{align}
\qquad [  \Omega,  X\otimes X]  =   [\Omega,  Y\otimes Y]     =  [\Omega, Z\otimes Z]  =  0 \, ,
\end{align}
and
\begin{align}
[\sigma_A,X]  =  [\sigma_A, Y]  =  [\sigma_A  ,  Z  ] =  0 \, ,
\end{align}
meaning that the test is covariant under the action of the Pauli group.

 It is  interesting to observe that for odd $N$, the set of states (\ref{eq-states})  is {\em not}  invariant under the action of the Pauli group.  In other words, the symmetries of the test are larger than the symmetries of the original set of states that the test is designed to probe.     This example illustrates the usefulness of our unified approach, in which the high-level structure of the benchmark (the operators $\Omega$ and $\sigma_A$) reveals symmetries that are not visible at the level of the original task that the device was meant to perform.

It is also important to stress that the  operators $\Omega$ and $\sigma_A$ are independent of $N$.  This means that  tests with different numbers of input states are equivalent in terms of score and probability of success.  In practice, this means that one can test the fidelity over {\em all the pure states on the equator}, by actually testing only the three states defined by Eq. (\ref{eq-states}) with $N=3$.
Alternatively, one can devise an equivalent test consisting in the preparation of the two-qubit maximally entangled state $|\Phi^+\>$, followed by the measurement of the two-qubit observable
\begin{align}
\nonumber \qquad O & =   2  \,      \Omega^{T_A}   \\
\nonumber &     =  \frac  1{2 } \,   \Big(   \ketbra  1 \otimes \ketbra 0  +    \ketbra 0 \otimes \ketbra 1  \\
 \nonumber   &  \qquad  \qquad    +   2  |\Phi^+\>\<\Phi^+|  \Big) \\
 &  = \frac 12  \,  \Big (  |\Psi^+\>\<\Psi^+  |   +  |\Psi^-\>\<\Psi^-|    +  2  \,  |\Phi^+\>\<\Phi^+|     \Big) \, ,
\end{align}
with $|\Psi^-\>  =  (|0\>\otimes |1\> -   |1\>\otimes |0\> )/\sqrt 2$.   The expectation value of the observable $O$ can be measured through a Bell measurement, or, indirectly, using the relation $O  =    ( I\otimes I  +  O_1  -  O_2)/2$, where $O_1$ and $O_2$ are the observables
\begin{align}
\qquad \nonumber O_1   =  |\Phi^+\>\<\Phi^+| \qquad {\rm and}  \qquad O_2  =   |\Phi^-\>\<\Phi^-|  \, ,
\end{align}
with  $|\Phi^-\>  =  (|0\>\otimes |0\> -   |1\>\otimes |1\> )/\sqrt 2$.     Using the above relation, the average fidelity over all pure states on the equator can be evaluated as
\begin{align}
F  =    \frac {1   +  \<  O_1\>  -  \<  O_2\>}2  \, ,
\end{align}
 where $\<O_1\>$ and $\<O_2\>$ are the expectation values of $O_1$ and $O_2$, respectively.
Also in this case, we can see the advantage of a more high-level formulation of the problem, which allowed us to find a setup that measures the average fidelity over all pure states on the equator by actually performing Bell measurements.

Now, we have seen that the fidelity test for the states  (\ref{eq-states}) is covariant under the action of the Pauli group, which is irreducible on the input system.  Since the performance operator is PPT,  Lemma \ref{lem:irr} yields the benchmark  expression
 \begin{align}
\nonumber ~\qquad S^{\rm(det)}_{\rm M\& P} & = 2 \,   \Lambda^{\otimes} ( \Omega )     \\
 \nonumber  &  =   2\,  \,   \sup_{\||\alpha\>\|  = 1 \, ,  \| |\beta\> \|=1} \,  \Tr \Big[\, \Big(\ketbra{\alpha} \otimes \ketbra{\beta} \Big)\,   \Omega \Big] \, ,
\end{align}
where the supremum can be evaluated using the Bloch representation
 \begin{align}
 \nonumber
 \ketbra{\alpha}   &=  \frac {  I  +    m_x \,    +  m_y\,   Y  +    m_z\,   Z }2  \\
 \ketbra{\beta}   &=  \frac {  I  +    n_x \,    X  +  n_y\,   Y  +    n_z\,  Z }2  \, .
    \end{align}
   Explicitly, we obtain
   \begin{align}
\nonumber   \Lambda^{\otimes 2}   (\Omega)  & =  \sup_{\st m, \st n}  ~\left\{\frac{4  +  2   (m_x  n_x  +  m_y n_y  )}{16}\right\}\\
  &  =  \frac 38\, ,
   \end{align}
so that the benchmark value is
\begin{align}
S^{\rm(det)}_{\rm M\& P}    =  \frac 34 \, .
\end{align}
Thanks to the symmetry of the problem, we know that the fidelity test is fully black box. Hence, every experimental fidelity above the classical threshold $3/4$ indicates a quantum advantage over arbitrary measure-and-prepare strategies, even including strategies that postselect on a subset of outcomes with arbitrary small probability.

\end{enumerate}

\subsection{The optimal measure-and-prepare strategy}

We have seen that tests that are covariant with respect to an irreducible representation are fully black box.   In this case, it is also possible to find an explicit expression for the  measure-and-prepare channel that achieves the benchmark.

\begin{lem} \label{lem:sym}
Let $\map T$ be a test for deterministic devices with PPT performance operator $\Omega$. Suppose that the input system $A$ has  dimension $d<\infty$.   If $\map T$  is covariant under the action of $\grp G$ and if and the group representation acting  on  $A$ is irreducible, then  the optimal measure-and-prepare channel is
\begin{align}
\map{C}(\rho) = \int \d g\,   \Tr[P_g \rho] ~ \ketbra{\psi_g}
\end{align}
where $\{P_g\}_{g \in G}$ is  the POVM defined  by
\begin{align}  P_g =    d \,  U_g      \ketbra  \phi U_g^\dag \, ,
\end{align}
for some unit vector $|\phi\>$, and the  output states $\{\ket{\psi_g}\}_{g \in G}$ have the form
\begin{align}
\ket{\psi_g} = U'_g \, \ket{\psi} \, ,
\end{align}
for another fixed unit vector $\ket\psi$.
\end{lem}

\Proof{} Consider a generic measure-and-prepare strategy with measurement of  the POVM $\{P_i\}$ and preparation of the states $\{\rho_i\}$.   The corresponding channel, denoted by $\map C$, is
\begin{align}
\map C(\rho)  =  \sum_i  \, \Tr  [  P_i  \, \rho] \,  \rho_i  \,,
\end{align}
and its \Jami{} operator is
\begin{align}
C  =   \sum_i  \rho_i\otimes P_i  \, .
\end{align}
Hence, the score is
\begin{align}
S^{\rm(det)}   (\map C):=& \sum_i \Tr\left[( \rho_i\otimes P_i) \, \Omega\right]\label{eq:det_fidelity_P}
\end{align}
 Without loss of generality, we assume that each operator $P_i$ is rank-one, because one can always split a non-rank-one operator into the sum of rank-one operators, without affecting the total score.  Likewise, we assume that each state $\rho_i$ is pure, because one can always include the randomization of pure states by adding dummy outcomes to the measurement, and use these outcomes to randomize over pure states.

Now, since the test is covariant, we have the equality
\begin{align}
\Omega = & \int \d g \,  (U'_g\otimes U_g) \Omega (U'_g\otimes U_g)^\dagger
\end{align}
Inserting this relation into  \Eq{eq:det_fidelity_P}, we obtain
\begin{align}
& S^{\rm(det)}(\map C) \nonumber\\  & =\sum_i \Tr\left[(\rho_i\otimes P_i )  \,  \int \d g \,  (U'_g\otimes U_g) \Omega (U'_g\otimes U_g)^\dagger  \right] \nonumber\\
\nonumber  &= \sum_i \int \d g \,    \Tr\left[ (  U_g^{\prime \,\dag}  \rho_i U'_g\otimes U_g^\dagger P_i U_g )~\Omega\right] \\
&    =  \sum_i  p_i  \,  \left(  \int \d g  \,      \Tr\Big[\Big( \rho^{(i)}_g \otimes   P^{(i)}_g \Big)   ~ \Omega\Big]\right) \, ,\label{eq:move_U}
\end{align}
having defined
\begin{align}
\nonumber p_i   &:=  \frac  {\Tr [P_i]}d \\
\nonumber \rho_g^{(i)}  & : =  U_g \, \rho_i \, U_g^\dag \\
P_g^{(i)}  &  :=  \frac{ d }{\Tr[  P_i]}  ~    U_g \, P_i \, U_g^\dag \, .
  \end{align}
Now, it is easy to check that
\begin{enumerate}
\item the numbers $\{p_i\}$ form a probability distribution
\item for every given  $i$, the operators $\{  P_g^{(i)}\}$ form a POVM, normalized as
\begin{align}
\int \d g\,  P_g^{(i)}  =  I_A \, .
\end{align}
\end{enumerate}
For given $i$, the POVM  $\{  P_g^{(i)}\}$ and the states $  \{\rho_g^{(i)}\}$ define a measure-and-prepare channel   $\map C_i$, with
\begin{align}\label{ci}
C_i (\rho)  = \int  \d g\,   \Tr\left [  P_g^{(i)}\,  \rho\right]  ~   \rho_g^{(i)} \, .
\end{align}
With this notation, Eq. (\ref{eq:move_U}) can be rewritten as
\begin{align}
\nonumber S^{\rm(det)}   (\map C)   &  =  \sum_i \, p_i  \,  S^{\rm(det)}   (\map C_i) \\
  &  \le \max_i  S^{\rm(det)}   (\map C_i)  \, .
\end{align}
Hence, the maximum score must be achieved by a measure-and-prepare channel $\map C_i$  of the form (\ref{ci}). Since by construction the POVM operators $P_g^{(i)}$ are rank-one, and sine the states   $\rho_g^{(i)}$ are pure, this concludes the proof. \qed

\end{document}